\documentclass[12pt,nohyper,letterpaper]{JHEP3}         % For preprint
\usepackage{amssymb,amsfonts}
\usepackage{epsfig}
\usepackage{amsmath}

\def\be{\begin{eqnarray}}
\def\ee{\end{eqnarray}}
\newcommand{\nn}{\nonumber}
\newcommand\para{\paragraph{}}
\newcommand{\ft}[2]{{\textstyle\frac{#1}{#2}}}
\newcommand{\eqn}[1]{(\ref{#1})}

\def\Dslash{\,\,{\raise.15ex\hbox{/}\mkern-12mu D}}
\def\Dbarslash{\,\,{\raise.15ex\hbox{/}\mkern-12mu {\bar D}}}
\def\delslash{\,\,{\raise.15ex\hbox{/}\mkern-9mu \partial}}
\def\delbarslash{\,\,{\raise.15ex\hbox{/}\mkern-9mu {\bar\partial}}}
\def\pslash{\,\,{\raise.15ex\hbox{/}\mkern-9mu p}}
\def\calDslash{\,\,{\raise.15ex\hbox{/}\mkern-12mu {\cal D}}}
\newcommand{\sign}{{\rm sign}}

\newcommand{\RN}{Reissner-Nordstr\"om\ }

\newcommand{\zstar}{z^\star}

\def\lae{\mathrel{\mathop{\smash{\lower .5 ex \hbox{$\stackrel<\sim$}}}}}
\def\lae{\mathrel{\mathop{\smash{\lower .5 ex \hbox{$\stackrel>\sim$}}}}}

\newcommand\silly{\bar{\Psi}\!\!\!\stackrel{\ \, \leftrightarrow}{\delslash}\!\Psi}

\title{A Gapless Hard Wall: Magnetic Catalysis in Bulk and Boundary}
\author{Stefano Bolognesi${}^1$, Jo\~{a}o N. Laia${}^2$, 
David Tong${}^{2}$ and Kenny Wong${}^2$ \\

${}^1$Racah Institute of Physics, The Hebrew University of Jerusalem, 91904, Israel \\ 
${}^2$Department of Applied Mathematics and Theoretical Physics, \\
University of Cambridge, UK\\
\\ {\ } \\
{\tt stefanbolo@gmail.com, j.laia, d.tong, k.wong@damtp.cam.ac.uk}
}

\abstract{We study various aspects of fermions and their chiral condensates, both in the bulk of AdS${}_4$ spacetime and in the dual boundary theory. For the most part, we focus on a geometry with an infra-red hard wall. We show that, contrary to common lore,  there exist boundary conditions in which the hard wall gives rise to a discrete, but gapless, fermionic spectrum. In such a setting, the presence of a magnetic field induces a bulk fermion condensate  which spontaneously breaks CP invariance. We develop the holographic dictionary between composite operators and show that this bulk condensate has the interpretation of boundary magnetic catalysis involving a double-trace operator. Finally, we explain how one can replace the hard wall with bulk magnetic monopoles. In such a framework, magnetic catalysis can be viewed as a consequence of the Callan-Rubakov effect.}

\begin{document}
\pagestyle{plain} \setcounter{page}{1}
\newcounter{bean}
\baselineskip16pt

\section{Introduction and Summary}

%Niel's Bohr once said that the opposite of  a deep truth is also a deep truth. In this paper we show that the holographic dual of magnetic catalysis is also magnetic catalysis. 

\para

The holographic hard wall is an anti-de Sitter geometry which terminates abruptly in the infra-red. It provides a  simple, yet robust,  caricature of holographic duals for theories with a gapped, discrete spectrum of states. Originally introduced to model confining gauge theories \cite{wall}, more recently the hard-wall geometry has been utilised to provide a holographic dual of the Fermi liquid state \cite{sachdev,mcgreevy}.

\para
The purpose of this paper is to study a number of inter-related aspects concerning fermions in the hard wall background. There are four punchlines which we summarise below.

\paragraph{A: A Gapless Hard wall}

\para
The holographic hard wall is synonymous with a mass gap in the boundary theory. However, at least for fermions, this need not necessarily be the case. The role of the hard wall is to make the spectrum discrete; it does not fix the mass of the lowest lying state.

\para
 In Section 2, we construct a one-parameter family of infra-red boundary conditions for fermions in an AdS hard wall geometry. As the boundary conditions vary, so too does the spectrum. We show that for one particular choice of boundary conditions, the fermion spectrum in the boundary theory is discrete, but gapless.

\paragraph{B: Holographic Magnetic Catalysis} 

\para
The phenomenon of  magnetic catalysis describes the formation of a fermion condensate due to the presence of a background magnetic field, spontaneously breaking a chiral symmetry \cite{magcat0}-\cite{magcat3}. 

\para
Unlike more familiar examples of chiral symmetry breaking, magnetic catalysis takes place at weak coupling.
In $d=2+1$ dimensions, the magnetic field induces a condensate for a massless Dirac fermion $\Psi$, 
\be \langle \bar{\Psi}\Psi\rangle = \pm \frac{B}{4\pi}\label{simple}\ee
The condensate is odd under parity P and even under charge conjugation C. However, the magnetic field itself is odd under both P and C. The net result is that the condensate spontaneously breaks the CP symmetry under which the magnetic field is invariant.

\para
On dimensional grounds, \eqn{simple} holds only at weak coupling where the condensate takes its canonical dimension, $[\bar{\Psi}\Psi]=[B]=2$. One can ask: what replaces \eqn{simple} in strongly coupled regimes where  $\bar{\Psi}\Psi$ acquires a large anomalous dimension?

\para
There have been a number of holographic studies of magnetic catalysis in an attempt to answer this question. The most common approach is to model the fermi bilinear as a scalar field in the bulk \cite{holomagcat1}-\cite{holomagcat6}. The well-developed mean-field machinery of scalar symmetry breaking in the bulk can then be brought to bear on the problem. 

\para
A different approach was taken in \cite{magcat}; the boundary fermion $\Psi$ was mapped directly to a bulk fermion $\psi$. Magnetic catalysis in the boundary was argued to be dual to magnetic catalysis in the bulk\footnote{The opposite of a deep truth is, of course, also a deep truth \cite{bohr}.}.

\para
The calculations in \cite{magcat} focussed primarily on the bulk fermions. It was shown that in the hard wall geometry, a bulk condensate $\langle\bar{\psi}\Gamma^5\psi\rangle$ is induced by a background magnetic field. However, a number of threads were left hanging. Firstly, for technical reasons, the bulk condensate could only be computed for massless bulk fermions. Secondly, although strong evidence was presented for the relationship between bulk and boundary condensates, no proof was given.

\para
In Section 3, we tie these threads. We show that  magnetic catalysis indeed takes place in the $d=2+1$ dimensional boundary theory, but only when the spectrum of fermions is both discrete and gapless. This is precisely the spectrum that arises from the gapless hard wall.  For a fermion of dimension $\Delta$, the weak coupling result \eqn{simple} is replaced by
\be  \langle \bar{\Psi}\Psi\rangle \sim  \pm \frac{B}{4\pi z^{\star\, 2-2\Delta}}\label{notsimple}\ee
where $1/\zstar$ is the gap induced by the hard wall between the lowest, massless state and the next state. (Holographically, $\zstar$ is the position of the hard wall). 

\paragraph{C: Composite Operators in AdS}

\para
The punchline of holography is that the  boundary theory knows everything that happens in the bulk. For example, if a fermion condensate forms in the bulk, the boundary theory knows. But how?

\para
For fermions in AdS${}_4$, it was suggested in \cite{magcat} that one can extend the holographic dictionary to embrace composite operators in both bulk and boundary.   Specifically, 

\be \bar{\psi}\Gamma^5\psi\ \longleftrightarrow \ \bar{\Psi}\Psi\nn\ee
This relationship is meant in the usual sense of AdS/CFT; by examining suitable fall-offs of the bulk composite operator $\bar{\psi}\Gamma^5\psi$, we can extract both the source and response of the boundary double trace operator $\bar{\Psi}\Psi$. This proposal was motivated by the observation that a $\langle\bar{\psi}\Gamma^5\psi\rangle$ condensate forms in the bulk of AdS$_4$ when \eqn{notsimple} forms on the boundary.  Moreover, it was shown in \cite{magcat} that this dictionary is consistent with the proposal of \cite{witten,berkooz} for implementing  sources for double trace operators through the use of mixed boundary conditions.  

\para
In Section 4, we confirm this proposal. We compute both the bulk condensate $\bar{\psi}\Gamma^5\psi$ as  a non-trivial function of our one-parameter family of hard wall boundary conditions and show how one can extract the boundary condensate $\bar{\Psi}\Psi$.

\para
Furthermore, we show that a similar relationship holds between the bulk and boundary composite operators
\be \bar{\psi}\psi\ \longleftrightarrow\ \bar{\Psi}\!\!\!\stackrel{\  \leftrightarrow}{\delslash}\!\Psi\nn\ee
It would certainly be interesting to see whether this kind relationship between composite operators in the bulk and boundary can be extended to other situations.

\paragraph{D: The Holographic Callan-Rubakov Effect}

\para
The general discussion of magnetic catalysis in a hard wall background suffers from a flaw: the hard wall must be magnetically charged. While there is no problem putting this in by hand, it is not obvious how this scenario may arise in any smooth geometry.

\para
In Section 5 of this paper, we provide a non-geometric alternative to the hard wall: a domain wall,  constructed out of non-Abelian gauge fields and scalars which lives in the full Poincar\'e patch of AdS. This domain wall can be viewed as knitted together out of 't Hooft-Polyakov monopoles and goes by the name of a {\it monopole wall}. It has the property that it spits out an approximately  homogeneous magnetic field towards the boundary of AdS.   Various aspects of the monopole wall in AdS were studied in \cite{monoads}. 

\para
The monopole wall is not a brutal truncation of AdS; it is merely a domain wall in the AdS geometry.  This means that  most excitations can pass through the monopole wall and it does not impose the same boundary conditions as the hard wall geometry. However, there is one exception: fermions lying in the lowest Landau level of the monopole wall are trapped in the UV end of the geometry. For them, the monopole wall acts as a reflecting boundary condition, entirely analogous to the hard wall geometry.

\para
In Section 5, we will show that much of the earlier discussion of this paper carries over in a natural way  to the monopole wall background.  Specifically, we will find that the bulk $\theta$-angle for non-Abelian gauge fields provides the one-parameter family of  IR boundary conditions on the monopole wall that we introduce in Section 2. 

\para
Moreover, the trapping of the lowest Landau-levels is entirely sufficient to induce the magnetic catalysis described in Section 3. In fact, within this framework, magnetic catalysis reduces to a well studied phenomenon: it is the bulk Callan-Rubakov effect \cite{rubakov,callan1,callan2}.

\para
Throughout the bulk of the paper we work in AdS${}_4$. A summary of the analogous results in AdS${}_5$ is provided in Appendix \ref{app5}, although the most interesting phenomena -- magnetic catalysis -- does not occur in this case due to a softening of infra-red effects. The more unsightly calculations of two-point functions and condensates have been relegated to the boundary, i.e.  appendices \ref{magwallapp} and \ref{hideapp}.

\section{Fermions in a Hard Wall Geometry}
\label{sec:2}

 The hard wall geometry is simply the Poincar\'e patch of AdS${}_4$,  truncated in the infra-red. We set the AdS radius to unity and work with coordinates 
 \be
ds^{2} = \frac{\eta_{ij}d x^i dx^j + dz^{2}}{z^{2}} \equiv \frac{-(dx^{0})^{2} + (dx^{1})^{2} + (dx^{2})^{2} + dz^{2}}{z^{2}}
\ee
with the radial coordinate $z$ restricted to lie in 
\be z\in [0,z^\star]\nn\ee
The UV boundary lies at $z=0$; the IR hard-wall lies at $z=\zstar$. The purpose of this section is to study the dynamics of a Dirac spinor, $\psi$ in the hard wall background.

\subsection{A circle of IR boundary conditions}
\label{sub:2.1}

The bulk  dynamics of the spinor $\psi$ of mass $m$ is governed by the standard Dirac action, 
 \begin{equation}
S_{\text{bulk}} = \int dz d^{3} x \sqrt{-g}\ \bar{\psi} \left( \frac{1}{2} \left[ e_{a}^{\mu}\, \Gamma^{a} \stackrel{\rightarrow}{D_\mu}
 - \stackrel{\leftarrow}{D_\mu} e_{a}^{\mu}\, \Gamma^{a} \right] - m  \right) \psi\label{sbulk}
\end{equation}
where the (inverse) vierbein is $e_{a}^{\mu} = z \delta_{a}^{\mu}$. Our chosen basis of four-dimensional gamma matrices is\footnote{The chiral gamma matrix $\Gamma^{5} \equiv i \Gamma^{z} \Gamma^{0} \Gamma^{1} \Gamma^{2} $ obeys $ (\Gamma^5)^2 = 1$. The hermiticity conditions  $\Gamma^{0} \Gamma^{a\,\dagger}  \Gamma^{0} = \Gamma^{a}$ hold. Conjugate spinors are defined by $\bar{\psi} = i\psi^{\dagger} \Gamma^{0} $; with this definition, $\bar{\psi} \psi$ is real and $\bar{\psi} \Gamma^5 \psi$ is imaginary}. A convenient representation is 
 \begin{equation}
\Gamma^{z} = \left( \begin{array}{cc} 1 & 0 \\ 0 & -1 \end{array} \right) \qquad
\Gamma^{i} = \left( \begin{array}{cc} 0 & \gamma^{i} \\ \gamma^{i} & 0 \end{array} \right) \qquad
\Gamma^{5} = \left( \begin{array}{cc} 0 & i \\ -i & 0 \end{array} \right), \label{rep}
\end{equation} 
where $\gamma^{0} = -i \sigma^{3}$, $\gamma^{1} = \sigma^{1}$ and $\gamma^{2} = \sigma^{2}$. These $\gamma^i$ furnish a representation of the $d=2+1$ boundary theory gamma matrices. 

\para
The variational principle leads directly to the Dirac equation which, substituting for the spin connection $\omega_{\mu}^{a b} = {z}^{-1} \left( \eta^{a z} \delta_{\mu}^{b}- \delta_{\mu}^{a} \eta^{b z} \right) $, becomes
\be
\left(e_{a}^{\mu} \Gamma^{a} D_{\mu} - m \right) \psi = \left( z \Gamma^{a} \partial_{a} - \tfrac{3}{2} \Gamma^{z} - m \right) \psi = 0.
\label{dirac}\ee
The bulk Dirac action action alone is not sufficient. Its variation results in boundary terms at both $z=0$ and $z=\zstar$ which, left unchecked, would fix all components of the spinor to vanish on both boundaries. 
%
%Varying the bulk action with respect to $\psi$ and $\bar{\psi}$, we obtain
%
% \begin{eqnarray*}
%\delta S_{\text{bulk}} & = &  \int dz d^3 x \sqrt{-g} \left( \delta \bar{\psi} \left( e_a^\mu \Gamma^a \overrightarrow{D_{\mu}} - m  \right) \psi + \bar{\psi} \left( - \overleftarrow{D_{\mu}} e_a^\mu \Gamma^a  - m  \right) \delta \psi \right) \\ 
%&& - \tfrac{1}{2} \int_{z = 0} d^3 x \sqrt{- h} \left( \delta \bar{\psi}^+ \psi^- - \delta \bar{\psi}^- \psi^+ + \bar{\psi}^- \delta \psi^+ - \bar{\psi}^+ \delta \psi^-  \right) \\
%&& + \tfrac{1}{2} \int_{z = z^\star} d^3 x \sqrt{- h} \left( \delta \bar{\psi}^+ \psi^- - \delta \bar{\psi}^- \psi^+ + \bar{\psi}^- \delta \psi^+ - \bar{\psi}^+ \delta \psi^-  \right),
%\end{eqnarray*}
%
This is overkill. The Dirac equation is a first order system, requiring us to fix half of the spinor components at the UV boundary \cite{henneaux,iqballiu}. (See \cite{laia} for a recent summary, where non-relativistic boundary conditions were also considered). Another half are fixed on the IR wall. The additional richness of fermions in the hard wall background comes from the interplay between which halves we choose to fix. 

\para
The choice of which components to fix on each boundary is determined by augmenting the bulk Dirac action with 
 the boundary term $S_{\rm bdy} = S_{\rm UV}+S_{\rm IR}$. We deal first with the UV boundary conditions at $z=0$. 
 
\para
To describe the boundary conditions, it is useful to project the spinor onto two subspaces, 
 \be \psi^\pm  = P_\pm \psi\ \ \ \ ,\ \ \ \ \  P_\pm = \frac{1}{2}\left(1\pm\Gamma^z\right)\nn\ee
 At the UV boundary, $z=0$, we will choose to fix the components of $\psi^-$. In fact, we can only legally do this for certain values of the mass because the fall-off of $\psi^+$ is only normalizable for $m>-1/2$. This means that if we want to allow the asymptotic value of $\psi^+$ to change dynamically (and we do) then we must restrict attention to this range of masses\footnote{There is no loss of generality here. For $m\leq -1/2$, we can exchange the role of $\psi_+$ and $\psi_-$, together with an appropriate redefinition of the IR boundary conditions.}. When $m>-1/2$, the leading  fall-offs of the spinor are given by
 \be \psi \rightarrow z^{3/2-m}\psi^-_0 + z^{3/2+m}\psi^+_0+\ldots \label{itfalls}\ee
 We will choose to fix $\psi^-_0$ as $z\rightarrow 0$.

\para
Imposing this boundary condition can be subsumed in the variational principle by the addition of the boundary term \cite{iqballiu},
\begin{equation}
S_\text{{UV}} = - \frac{1}{2} \int _{z = 0} d^{3}x \sqrt{-h}\ \bar{\psi} \psi\label{suv}
\end{equation}
%Varying this action, we obtain 
%
%\begin{equation*}
%\delta S_{\text{UV}} = -\tfrac{1}{2} \int _{z = 0} d^{3}x \sqrt{-h} \left( \delta \bar{\psi}^+ \psi^- + \delta \bar{\psi}^- \psi^+ + \bar{\psi}^- %\delta \psi^+ + \bar{\psi}^+ \delta \psi^-  \right)
%\end{equation*} 
%
It is a simple exercise to check that the variation of  $S_{\rm bulk}+S_{\rm UV}$ vanishes only if we hold  $\psi^-$ fixed on the boundary.  For $m\geq 0$, this choice is referred to as {\it standard quantization}; for $-1/2<m<0$, it is referred to as {\it alternative quantization}.

\para
Translated via the AdS/CFT correspondence, $\psi^-_0$ is interpreted as  the source for the 2-component Dirac spinor operator $\Psi$ in the $d=2+1$ dimensional boundary theory. Meanwhile, $\psi^{+}_0$ plays the part of the response, setting the conformal scaling dimension of the boundary operator to be $\Delta[\Psi]=\frac{3}{2}+m$. 

%In the representation \eqref{rep}, $\psi = \left( \begin{array}{c} \psi^{+} \\ \psi^{-} \end{array} \right) $ and $\bar{\psi} = (\bar{\psi}^{-}, \bar{\psi}^{+})$. The \emph{top-right} blocks of the bulk gamma matrices couple together the sources $\psi^{-}$ and $\bar{\psi}^{-}$. These top-right blocks, the $\gamma^{i}$s, are inherited by the boundary theory.

\para
Our choice for the boundary conditions on the infra-red hard wall at $z=\zstar$ is more varied. We may pick any Hermitian, Lorentz invariant action so long as the resulting boundary conditions fix half of the spinor components. We will explore a one-parameter family of boundary conditions, labelled by a chiral angle  $\theta \in (-\pi , \pi]$. These are defined by the boundary term,
 \begin{equation}
S_{\text{IR}} = \frac{1}{2} \int _{z = z^{\star}}  d^{3}x \sqrt{-h} \ \bar{\psi}  \exp \left( i \theta \Gamma^5 \right) \psi
\label{sir}\end{equation}
%
%Varying it, we find 
%
%\begin{multline*}
%\delta S_{\text{IR}} = \tfrac{1}{2} \int _{z = z^\star} d^{3}x \sqrt{-h} \left( \cos\theta \left( \delta \bar{\psi}^+ \psi^- + \delta \bar{\psi}^- \psi^+  + \bar{\psi}^- \delta \psi^+ + \bar{\psi}^+ \delta \psi^- \right) \right. \\
%\left. + i \sin\theta \left( \delta \bar{\psi}^+ \Gamma^5 \psi^+ + \delta \bar{\psi}^- \Gamma^5 \psi^- + \bar{\psi}^+ \Gamma^5 \delta \psi^+ + \bar{\psi}^- \Gamma^5 \delta \psi^-  \right)\right), 
%\end{multline*} 
%
The variational principle now requires that a linear combination of $\psi^+$ and $\psi^-$ vanish on the wall\footnote{The IR boundary conditions arising from \eqn{sir} are similar to the circle of UV boundary conditions considered previously for massless fermions \cite{porrati,redi}. Indeed, in \cite{magcat}, which discussed only massless bulk fermions, the IR boundary conditions were fixed and the UV boundary conditions rotated.}
\begin{equation} 
\cos\left(\tfrac{\theta}{2}\right)
 \psi^{-} + i \sin \left(\tfrac{\theta}{2}\right) \Gamma^{5} \psi^{+}  = 0 \qquad  \text{at $z=\zstar$} \label{wallbc}
\end{equation}
As we will explore in more detail shortly, each choice of $\theta$ corresponds to a distinct boundary theory with a distinct spectrum. The choice usually employed in hard wall geometries \cite{sachdev} corresponds to $\theta=\pi$.

\subsection{Discrete Symmetries: P, C and CP}

Parity and charge conjugation will be central to the later part of our stories. It will prove useful to spell out how these discrete symmetries are affected by the IR boundary condition \eqn{wallbc}.

\para
We will define parity in both the bulk and boundary theories as a mirror reflection in just a single spatial direction which we take to be $x^1$.  The effect of parity on the bulk spinor is then $\psi_{P} = - i \Gamma^1 \Gamma^5 \psi$ and the fermion bilinears transform as
\begin{equation}
(\bar{\psi} \psi)_{P} = + \bar{\psi} \psi \qquad
(\bar{\psi} \Gamma^{5} \psi)_{P} = - \bar{\psi} \Gamma^{5} \psi.
\end{equation}

\para
We  implement charge conjugation by $ \psi_{C} = C \bar{\psi}^{\,T}$, where the charge conjugation matrix $C$ satisfies $C \Gamma^{a T} C^{-1} = -\Gamma^{a}$ and impose $C^{\dagger} = C^{-1}$ and $C^T=-C$.
%
% We have $\bar{\psi}_{C} = - \psi^T C^{-1}$. If we further impose $C^{T} = -C$, then for any matrix $A$, $(\bar{\psi}A\psi)_C = \bar{\psi}CA^TC^{-1}\psi$. (It is important to remember that spinors are anticommuting objects.) In the representation \eqref{rep}, all of these conditions are realised by 
%
In the representation \eqn{rep}, we choose
\begin{equation*}
C = \left( \begin{array}{cc} 0 & \sigma^{2} \\ \sigma^{2} & 0 \end{array} \right).\end{equation*} 
The two fermion bilinears are then both even under charge conjugation,
\begin{equation}
(\bar{\psi} \psi)_{C} = + \bar{\psi} \psi \qquad
(\bar{\psi} \Gamma^{5} \psi)_{C} = +\bar{\psi} \Gamma^{5} \psi.
\end{equation}
Turning now to our infra-red hard wall, we see that for generic values of $\theta$, the boundary term \eqn{wallbc} breaks P and hence CP. There are just two exceptions: $\theta=0$ and $\theta=\pi$. In these two, special cases, both P and CP remain as symmetries of the theory.

\para
The action of the discrete symmetries on the boundary spinor $\Psi$ is inherited from the bulk transformation above.  Under parity, 
 $\Psi_{P} = \sigma^{1} \Psi$ and $\bar{\Psi}_{P} = - \bar{\Psi}\sigma^{1}$. Under charge conjugation,  $\Psi_{C} = \sigma^{2} \bar{\Psi}^{T}$ and $\bar{\Psi}_{C} = - \Psi^{T} \sigma^{2}$. The boundary bilinear transforms as
\be (\bar{\Psi}\Psi)_P=-\bar{\Psi}\Psi\ \ \ \ ,\ \ \ \ (\bar{\Psi}\Psi)_C=+\bar{\Psi}\Psi\nn\ee
These are the usual transformation properties for spinors in $d=2+1$. Notice in particular that $\bar{\Psi} \Psi$ possess the same discrete symmetry transformations as $i \bar{\psi} \Gamma^{5} \psi$. It is an innocent little observation, but an important one -- it foreshadows much of the story of Section \ref{sec:3}.

%\begin{equation}
%\left( \tfrac{1}{2} \bar{\Psi} (\gamma^i \overrightarrow{\partial_i} - \overleftarrow{\partial_i} \gamma^i) \Psi \right)_{P} = + \tfrac{1}{2} \bar{\Psi} (\gamma^i \overrightarrow{\partial_i} - \overleftarrow{\partial_i} \gamma^i) \Psi \qquad
%(\bar{\Psi}\Psi)_{P} = - \bar{\Psi}\Psi.
%\end{equation}
%
%\paragraph{} Under charge-conjugation, the boundary spinors inherit the transformations.\begin{equation}
%\left( \tfrac{1}{2} \bar{\Psi} (\gamma^i \overrightarrow{\partial_i} - \overleftarrow{\partial_i} \gamma^i) \Psi \right)_{C} = + \tfrac{1}{2} \bar{\Psi} (\gamma^i \overrightarrow{\partial_i} - \overleftarrow{\partial_i} \gamma^i) \Psi \qquad
%(\bar{\Psi}\Psi)_{C} = + \bar{\Psi}\Psi
%\end{equation}

\subsection{The Spectrum}
\label{sub:2.2}

Our first task is to examine how the spectrum of the theory varies with $\theta$. The general solution to the Dirac equation \eqn{dirac} is expressed in terms of Bessel functions
%\footnote{To verify this solution, it is necessary to apply the Bessel function identities $Y_\nu ' (u) = \frac{1}{2} \left( Y_{\nu - 1} (u) - Y_{\nu + 1} (u) \right)$ and $\nu Y_\nu (u) = \frac{u}{2} \left( Y_{\nu - 1} (u) + Y_{\nu + 1} (u) \right)$,  as well as the corresponding ones with $Y_\nu(u)$ replaced by $J_\nu(u)$.}
 \cite{henning,muck}
\begin{multline}
\psi(z,\vec{x}) = \int \frac{d^{3}k}{(2\pi)^{3}}e^{-i \vec{k} \cdot \vec{x}} z^{2} \left( Y_{m+\frac{1}{2}}(kz) a^{-}(\vec{k}) + Y_{m-\frac{1}{2}}(kz) a^{+}(\vec{k})  \right. \\
\left. + J_{m+\frac{1}{2}}(kz) b^{-}(\vec{k}) + J_{m-\frac{1}{2}}(kz) b^{+}(\vec{k}) \right), \label{soln}
\end{multline}
with the plus and minor polarisation spinors  related by 
\begin{equation*}
a^{+}(\vec{k}) = - i \vec{\Gamma} \cdot\hat{\vec{k}}\, a^{-}(\vec{k}) \qquad
b^{+}(\vec{k}) = - i \vec{\Gamma} \cdot \hat{\vec{k}}\, b^{-}(\vec{k}).
\end{equation*} 
Here the vector arrow denotes directions parallel to the boundary, including time. We define $k = (-\eta^{i j} k_{i} k_{j})^{1/2}$ and $\hat{\vec{k}} = \vec{k} / k$, so $( \vec{\Gamma} \cdot \hat {\vec{k}} )^2 = - 1$.\footnote{When $k_i$ is timelike, $k$ is real and positive. When $k_i$ is spacelike, $k$ may take either of two imaginary values. The case where $k_i$ is null is treated specially below.}

\para
The boundary conditions on the hard wall restrict the possible values of $k$ to a discrete set, corresponding to the spectrum of masses of excitations in the boundary dual. The UV boundary condition immediately sets the $a(\vec{k})$ polarisation spinors to zero. The IR boundary condition \eqref{wallbc} imposes a restriction on the $b(\vec{k})$ polarisation spinors, 
\begin{equation}
\left[\cos\left(\tfrac{\theta}{2}\right)\, J_{m+\frac{1}{2}}(kz^{\star}) + \sin\left( \tfrac{\theta}{2}\right)\,J_{m-\frac{1}{2}}(kz^{\star})\  \Gamma^{5} \vec{\Gamma} . \hat{\vec{k}}\, \right] b^{-}(\vec{k}) = 0, \label{bcond}
\end{equation}
which admits a non-zero $b^- (\vec{k})$ if and only if 
\begin{equation}
\frac{J_{m+\frac{1}{2}}(kz^{\star})}{J_{m-\frac{1}{2}}(kz^{\star})}= \pm \tan\tfrac{\theta}{2}. \label{spec}
\end{equation}
%
 %To prove that \eqref{spec} is a necessary condition, we premultiply \eqref{bcond} by $\cos\tfrac{\theta}{2} J_{m+\frac{1}{2}}(kz^{\star}) - \sin \tfrac{\theta}{2} \Gamma^{5} \vec{\Gamma} . \hat{\vec{k}}J_{m-\frac{1}{2}}(kz^{\star})$. 
Whenever \eqn{spec} holds,  any $b^-(\vec{k})$ in the kernel of the projection matrix $\frac{1}{2} (1 \pm \Gamma^5 \vec{\Gamma}\cdot \hat{\vec{k}})$ satisfies \eqref{bcond}.

\para
As expected for the hard wall geometry, there is a Kaluza-Klein tower of states. Every $k$ that obeys \eqref{spec} is the mass of an excitation in the spectrum of the dual boundary theory. 
When $m=0$, the condition \eqn{spec} simplifies. (Indeed, this massless limit will be a testing ground for later calculations). 
In this special case the Bessel functions reduce to friendly trigonometric functions and the spectrum is perfectly periodic. 
\begin{equation*}
k = \frac{1}{z^{\star}} \left( \pm \frac{\theta}{2} + \pi n \right) \qquad \qquad (m=0)\label{m0spec}
\end{equation*}
However, the spectrum is not qualitatively different for any value of the mass  $m > -\frac{1}{2}$ (i.e. for boundary fermions with conformal dimensions greater than the free value $1$). In Figure 1, we plot the spectrum as a function of $\theta$ for different values of the masses,

\begin{picture}(400,235)
\put(143,0){Figure 1. The spectrum}
\put(187,25){$m=0$}
\put(150,60){\line(1,0){100}}
\put(150,60){\vector(0,1){140}}
\put(200,45){$\theta$}
\put(150,45){$0$}
\put(245,45){$\pi$}
\put(140,180){$k$}
\qbezier(150,60)(200,66)(250,72)
\qbezier(250,72)(200,78)(150,84)
\qbezier(150,84)(200,90)(250,96)
\qbezier(250,96)(200,102)(150,108)
\qbezier(150,108)(200,114)(250,120)
\qbezier(250,120)(200,126)(150,132)
\qbezier(150,132)(200,138)(250,144)
\qbezier(250,144)(200,150)(150,156)
\qbezier(150,156)(200,162)(250,168)
\qbezier(250,168)(200,174)(150,180)
\qbezier(150,180)(175,183)(200,186)
\put(255,132){\vector(0,1){12}}
\put(255,132){\vector(0,-1){12}}
\put(258,129){$\tfrac{\pi}{z^{\star}}$}

\put(327,25){$m>0$}
\put(290,60){\line(1,0){100}}
\put(290,60){\vector(0,1){140}}
\put(340,45){$\theta$}
\put(290,45){$0$}
\put(385,45){$\pi$}
\put(280,180){$k$}
\qbezier(290,60)(340,79.9)(390,89.3)
\qbezier(390,89.3)(340,94.3)(290,99.1)
\qbezier(290,99.1)(340,106.6)(390,113.5)
\qbezier(390,113.5)(340,118.8)(290,124.3)
\qbezier(290,124.3)(340,131)(390,137.8)
\qbezier(390,137.8)(340,143.3)(290,148.8)
\qbezier(290,148.8)(340,155.3)(390,161.8)
\qbezier(390,161.8)(340,167.3)(290,173.8)
\qbezier(290,173.8)(340,179.5)(390,185.8)

\put(28,25){$-\tfrac{1}{2}<m<0$}
\put(10,60){\line(1,0){100}}
\put(10,60){\vector(0,1){140}}
\put(60,45){$\theta$}
\put(10,45){$0$}
\put(105,45){$\pi$}
\put(0,180){$k$}
\qbezier(10,60)(60,62.6)(110,67.2)
\qbezier(110,67.2)(60,74.1)(10,79.9)
\qbezier(10,79.9)(60,86.4)(110,92.2)
\qbezier(110,92.2)(60,98.4)(10,103.9)
\qbezier(10,103.9)(60,110.4)(110,116.2)
\qbezier(110,116.2)(60,122.4)(10,128.4)
\qbezier(10,128.4)(60,134.4)(110,140.2)
\qbezier(110,140.2)(60,146.2)(10,152.6)
\qbezier(10,152.6)(60,158.4)(110,164.4)
\qbezier(110,164.4)(60,170.4)(10,176.4)
\qbezier(10,176.4)(60,182.4)(110,188.4)
\end{picture}

\paragraph{} For any given mode, the $P^{\pm}$ ``chirality'' of the spinor rotates like a corkscrew whose axis is aligned along the radial direction. Both ends of the corkscrew must slot into the grooves set by the UV and IR boundary conditions. Every other rung in the Kaluza-Klein ladder add an extra full twist to the corkscrew. The distinction between odd- and even-numbered rungs is a distinction of handedness.

\subsubsection*{The Gapless Mode}

\paragraph{} As $\theta \to 0$ or $\pi$, pairs of modes coalesce into doubly degenerate modes (the grooves of the corkscrew are parallel or perpendicular; left- and right-handed corkscrews are equivalent as a result of restored P invariance). What is most important for our discussion however is that the $\theta = 0$ theory has a massless excitation, a lone mode with $k=0$. The $\theta = 0$ spectrum is discrete but {\it gapless.}

\para
The  $\theta = 0$ boundary condition kills $\psi^{-}$ both in the IR and the UV. But resulting the $k=0$ zero modes are ``chiral'' with respect to the projection operator $P^{\pm}$: the corkscrew is not a corkscrew any more -- it does not rotate and the  purely plus solution slots neatly into the grooves.
The solution \eqn{soln} breaks down in the limit of a null vector with $k^2=0$. However, it is not difficult to find its replacement:
\be
\psi(z, \vec{x}) = e^{-i\vec{k}\cdot\vec{x}}z^{\frac{3}{2}+m}b^{+}(\vec{k})
\ee
where $b^+(\vec{k})$ lies in the kernel of $i\vec{\Gamma}\cdot\vec{k}$ (which, since $k$ is null, necessarily has a zero eigenvalue).

\subsection{Boundary Correlation Functions}
\label{sub:2.3}

We now turn to the computation of the boundary correlation functions as a function of $\theta$. The bulk fermion $\psi$ is dual to a two-component spinor operator $\Psi$ in the $d=2+1$ boundary CFT. Extracting the boundary theory two-point function $\langle \bar{\Psi}_{\beta} (\vec{y}) \,\Psi_{\alpha} (\vec{x}) \rangle$ is a routine AdS/CFT calculation \cite{henning,muck}. The spectra derived in Section \ref{sub:2.2} dictate the form of the two-point function: every mode in the spectrum is realised as a pole in its momentum space representation, and the residues of the poles have exponential decays in Euclidean position space that are the signature of systems with discrete spectra.

\para
As usual, the generating function for correlation functions is provided by the on-shell spinor action in AdS: $S_{\rm AdS}=S_{\rm bulk}+S_{\rm UV} + S_{\rm IR}$. (The individual terms in the action were defined in equations \eqn{sbulk}, \eqn{suv} and \eqn{sir}). As described in Section \ref{sub:2.1}, the leading components $\psi_0^-$ near the boundary play the role of the source for $\Psi$ and the two-point function is given by,
%
%Following \cite{muck}, we define the source $\psi_{0}^{-} = \lim_{z \to 0} z^{-\frac{3}{2}+m} \psi^{-}$ and the response $\psi_{0}^{+} = \lim_{z \to 0} z^{-\frac{3}{2}-m} \psi^{+}$. The AdS/CFT correspondence equates the bulk partition function to the boundary generating functional. At strong coupling and large $N$, the bulk partition function can be approximated by the classical, on-shell AdS action.\begin{equation*}
%1 + i S_{\text{AdS}}(\psi_0^-, \bar{\psi}_0^-) \approx \exp \left( i S_{\text{AdS}}(\psi_0^-, \bar{\psi}_0^-) \right)  = \exp \langle i \int \left( \bar{\Psi} \psi_0^- + \bar{\psi}_0^- \Psi \right) \rangle
%\end{equation*} The two-point function of the boundary fermion $\Psi$ can then be calculated by taking functional derivatives of the AdS action with respect to the source $\psi_0^-$.
%
\be
\langle \bar{\Psi}_{\beta} (\vec{y}) \,\Psi_{\alpha} (\vec{x}) \rangle
= -i\left.\frac{\delta}{\delta \psi_{0 \beta}^{-}(\vec{y})} \,\frac{\delta}{\delta \bar{\psi}_{0 \alpha}^{-}(\vec{x})}\ S_{\text{AdS}}\right|_{{\psi_{0}^{-} = 0, \bar{\psi}_{0}^{-} = 0}}
\nn\ee
The Dirac equation clearly ensures $S_{\rm bulk}=0$ on-shell. Similarly, the IR boundary conditions ensures that $S_{\rm IR}=0$ on-shell. 
This leaves the UV boundary term \eqn{suv} as the only contribution to the on-shell action: it survives simply because we must relax the boundary condition to allow a non-zero source, setting $\psi_0^-=0$, only after taking the functional derivative. We therefore have
\be
S_{\text{AdS}} = -\frac{1}{2} \int d^{3}x \left( \bar{\psi}_{0}^{-}(\vec{x}) \psi_{0}^{+}(\vec{x}) + \bar{\psi}_{0}^{+}(\vec{x}) \psi_{0}^{-}(\vec{x})\right) \label{chocolate}
\ee
where the response $\psi^+_0$ is determined in terms of the source $\psi^-_0$ by the infra-red boundary conditions.  

\para
To determine this relationship, we first substitute the mode expansion \eqn{soln} into the infra-red boundary condition \eqn{wallbc} to arrive at  
a condition between the polarisation spinors  $a^{-}({k})$ and $b^{-}(\vec{k})$,
\begin{equation*}
b^{-}(\vec{k}) = \left[ M_{1}(k) + M_2(k)\,\Gamma^5 \, \vec{\Gamma} . \hat{\vec{k}}\, \right]a^{-}(\vec{k})
\end{equation*}
where  the coefficients  are given by
%
%\be
%M_{1}(k)
%& \equiv & \frac{1}{\Upsilon} \left( - \cos^{2}\left(\frac{\theta}{2}\right)) J_{m+\frac{1}{2}}(kz^{\star}) Y_{m+\frac{1}{2}}(kz^{\star}) + \sin^{2}\left(\frac{\theta}{2}\right) J_{m-\frac{1}{2}}(kz^{\star}) Y_{m-\frac{1}{2}}(kz^{\star}) \right)\nn\ee
%
%and 
%
%\be
%M_{2}(k)
%& \equiv & -\frac{1}{\Upsilon}\,\frac{\sin\theta}{\pi k\zstar}
%\label{coeff2}
%\ee
%with
%\be \Upsilon\equiv {\cos^{2}\left(\frac{\theta}{2}\right) J_{m+\frac{1}{2}}^{2}(kz^{\star}) -\sin^{2}\left(\frac{\theta}{2}\right) J_{m-\frac{1}{2}}^{2}(kz^{\star})}, \nn\ee
%
%
%
\begin{eqnarray}
M_{1}(k)
& \equiv & \frac{ - \cos^{2}(\tfrac{\theta}{2}) J_{m+\frac{1}{2}}(kz^{\star}) Y_{m+\frac{1}{2}}(kz^{\star}) + \sin^{2}(\tfrac{\theta}{2}) J_{m-\frac{1}{2}}(kz^{\star}) Y_{m-\frac{1}{2}}(kz^{\star})}
{\cos^{2}(\tfrac{\theta}{2}) J_{m+\frac{1}{2}}^{2}(kz^{\star}) -\sin^{2}(\tfrac{\theta}{2}) J_{m-\frac{1}{2}}^{2}(kz^{\star})}, \notag
\\
M_{2}(k)
& \equiv & -\frac{1}{\pi k\zstar}\, \frac{\sin \theta  }
{\cos^{2}(\tfrac{\theta}{2}) J_{m+\frac{1}{2}}^{2}(kz^{\star}) -\sin^{2}(\tfrac{\theta}{2}) J_{m-\frac{1}{2}}^{2}(kz^{\star})}.
\label{coeffs}
\end{eqnarray}
%
%The identity $J_{\nu+\frac{1}{2}}(u)Y_{\nu-\frac{1}{2}}(u)-Y_{\nu+\frac{1}{2}}(u)J_{\nu-\frac{1}{2}}(u) = 2 / \pi u $ is used to derive the expression for $M_2(k)$.
%
With this relation, together with the standard series expansions for the Bessel functions\footnote{
%We take the $O(z^{\frac{3}{2}-m})$ terms in $Y_{m+\frac{1}{2}}(kz)$ and $J_{m+\frac{1}{2}}(kz)$, and we take the $O(z^{\frac{3}{2}+m})$ terms in $Y_{m-\frac{1}{2}}(kz)$ and $J_{m-\frac{1}{2}}(kz)$. These terms are leading only when $-\tfrac{1}{2} < m < \tfrac{1}{2}$. 
For $m>1/2$, there is an additional divergent term that must be removed by holographic renormalization. It arises because $\psi^+$ has two different fall-offs in UV. Expanding in small $z$, $\psi^+$ has the asymptotic behaviour $\psi^+ \sim Az^{3/2+m}+Bz^{5/2-m}$. The coefficient $A$ is used to determine the boundary two-point function \eqn{corr}. When $m>1/2$, the coefficient $B$ gives rise to a contact term in the boundary two-point function which must be removed by holographic renormalization. See  \cite{iqballiu} for a discussion.}, we can to use \eqn{soln} to express  $\psi_{0}^{-}$ and $\psi_{0}^{+}$ in terms of the mode spinors. The source is given by
\be
\psi_0^-(\vec{x}) & = & - \int \frac{d^3k}{(2\pi)^3} e^{-i\vec{k}.\vec{x}}\left( \frac{k}{2}\right)^{-m-\frac{1}{2}} \frac{\Gamma(m+\tfrac{1}{2})}{\pi} a^-(\vec{k}) \nn\ee
while the response is 
\be
\psi_0^+(\vec{x}) & = & - \int \frac{d^3k}{(2\pi)^3} e^{-i\vec{k}.\vec{x}}\left( \frac{k}{2}\right)^{m-\frac{1}{2}} \nn \\
&& \qquad \times \left[ i \vec{\Gamma} . \hat{\vec{k}} \left(\frac{M_1(k)}{\Gamma(m + \tfrac{1}{2})} - \frac{\sin(\pi m) \Gamma(-m+ \tfrac{1}{2})}{\pi}\right) + i \Gamma^5 \frac{M_2(k)}{\Gamma(m + \tfrac{1}{2})} \right] a^-(\vec{k})
\nn\ee
With these results, we can achieve our aim of expressing the on-shell action entirely in terms of the source, 
\begin{equation}
S_{\rm AdS} = - \int d^{3} x d^{3} y \frac {d^{3} k} {(2 \pi)^{3}} \left( \frac{k}{2} \right) ^{2m} e^{-i \vec{k} \cdot(\vec{x}-\vec{y})}\, \bar{\psi}_{0}^{-} (\vec{x}) \left[ i \vec{\Gamma} . \hat{\vec{k}} N_{1} (k) +  i \Gamma^{5} N_{2} (k) \right] \psi_{0}^{-} (\vec{y}) \label{onshell}
\end{equation} 
where 
\begin{eqnarray}
N_{1} (k) & \equiv & \frac{\pi \cos^{2} (\tfrac{\theta}{2}) J_{m+\frac{1}{2}}(kz^{\star}) J_{-m-\frac{1}{2}}(kz^{\star}) + \pi \sin^{2} (\tfrac{\theta}{2}) J_{-m+\frac{1}{2}}(kz^{\star}) J_{m-\frac{1}{2}}(kz^{\star})   } {\cos (m \pi) \Gamma(m + \tfrac{1}{2})^{2} \left( \cos^{2} (\tfrac{\theta}{2}) J_{m+\frac{1}{2}}^{2}(kz^{\star}) - \sin^{2} (\tfrac{\theta}{2}) J_{m-\frac{1}{2}}^{2}(kz^{\star}) \right)}, \notag \\
N_{2} (k) & \equiv & -\frac{ \sin \theta  } { k\zstar \Gamma(m + \tfrac{1}{2})^{2} \left( \cos^{2} (\tfrac{\theta}{2}) J_{m+\frac{1}{2}}^{2}(kz^{\star}) - \sin^{2} (\tfrac{\theta}{2}) J_{m-\frac{1}{2}}^{2}(kz^{\star}) \right)}. \label{coeffs2}
\end{eqnarray}
Once functional derivatives are taken with respect to the sources, and the UV boundary condition is applied, we obtain our desired expression for the boundary two-point function in momentum space,
\begin{equation}
\langle \bar{\Psi}_{\beta} (\vec{y})\, \Psi_{\alpha} (\vec{x}) \rangle  = -i \int \frac {d^{3} k} {(2 \pi)^{3}} \left( \frac{k}{2} \right) ^{2m} e^{-i \vec{k} \cdot (\vec{x}-\vec{y})} \left( -  i \vec{\gamma}_{\alpha \beta}\cdot \hat{\vec{k}} \, N_{1} (k) + \delta_{\alpha \beta}\, N_{2} (k)  \right) \label{corr}
\end{equation} 
Reassuringly, $N_{1}(k)$ and $N_{2}(k)$ have single poles at every point in the spectrum. Equally reassuring is the observation that as $\zstar\rightarrow\infty$ and  the wall recedes way into the IR, our expression reproduces the conformal result  of \cite{henning,muck} for fermions with scaling dimension $\Delta[\Psi]=\frac{3}{2}+m$.  Note, in particular, that $N_2$ vanishes as $\zstar\rightarrow\infty$, leaving the  two-point function traceless.

\subsection{An Example: The Massless Bulk Fermion}
\label{subway}

To determine the two-point function \eqn{corr} in position space, we have a rather daunting Fourier transform to perform. As we now show, for massless bulk fermions with $m=0$, the Bessel functions simplify and the Fourier transformations become analytically calculable. As well as illustrating how the hard wall affects the IR physics, the resulting expressions will also be useful in subsequent sections. 

\para
For $m=0$, the simple poles of $N_{1}(k)$ and $N_{2}(k)$ are located at each  point in the spectrum \eqn{m0spec}. In Minkowski space, these lie neatly along the real $k$-axis. We can swing them onto the imaginary $k$-axis by Wick rotating $k_{0} \to i k_{0}$ according to the Feynman prescription. Let us redefine $k = \sqrt{\delta^{ij} k_{i} k_{j}}$. Performing the angular integrals in momentum space, we are left with the integral 
\begin{multline}
\langle \bar{\Psi}_{\beta} (\vec{0}) \Psi_{\alpha} (\vec{r}) \rangle  = -\frac{1}{4\pi^2} \left( \vec{\sigma}_{\alpha \beta} \cdot \hat{\vec{r}}\  \frac{\partial}{\partial r} \frac{1}{r} \Im \int_{-\infty}^{\infty} dk\ e^{ikr} \frac{\sinh 2kz^{\star}}{\cosh 2kz^{\star} - \cos \theta}  \right. \\ +  \left. \delta_{\alpha \beta} \frac{1}{r} \frac{\partial}{\partial r}\, \Re \int_{-\infty}^{\infty} dk\ e^{ikr} \frac{\sin\theta}{\cosh 2kz^{\star} - \cos \theta} \right)
\end{multline}
The Fourier transform\footnote{Fourier transforms are defined on a wider class of objects than normal Lebesgue integrals. The first of our two inverse Fourier transforms converges within the theory of distributions.} can be evaluated by completing the contour as in Figure 2, enclosing the infinite series of poles. Each of these poles leaves its mark on the correlator, with every residue contributing an  exponential fall-off in $r$, decaying as the Compton wavelength of the associated excitation.

\begin{picture}(400,160)
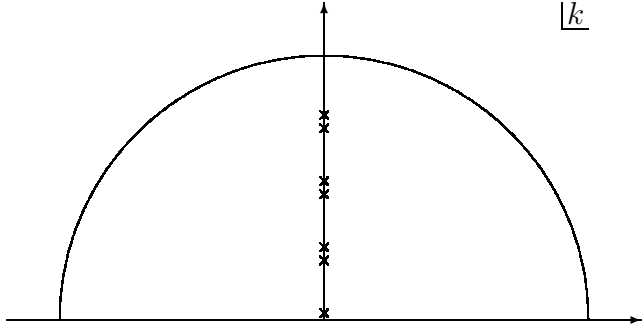
\put(50,0){Figure 2. The pole structure dictates the two-point function}
\put(200,20){\vector(0,1){120}}
\put(80,20){\vector(1,0){240}}

\qbezier(300.0,20.0)(299.9,24.4)(299.6,28.7)
\qbezier(299.6,28.7)(299.1,33.1)(298.5,37.4)
\qbezier(298.5,37.4)(297.6,41.6)(296.6,45.9)
\qbezier(296.6,45.9)(295.4,50.1)(294.0,54.2)
\qbezier(294.0,54.2)(292.4,58.3)(290.6,62.3)
\qbezier(290.6,62.3)(288.7,66.2)(286.6,70.0)
\qbezier(286.6,70.0)(284.3,73.7)(281.9,77.4)
\qbezier(281.9,77.4)(279.3,80.9)(276.6,84.3)
\qbezier(276.6,84.3)(273.7,87.6)(270.7,90.7)
\qbezier(270.7,90.7)(267.6,93.7)(264.3,96.6)
\qbezier(264.3,96.6)(260.9,99.3)(257.4,101.9)
\qbezier(257.4,101.9)(253.7,104.3)(250.0,106.6)
\qbezier(250.0,106.6)(246.2,108.7)(242.3,110.6)
\qbezier(242.3,110.6)(238.3,112.4)(234.2,114.0)
\qbezier(234.2,114.0)(230.1,115.4)(225.9,116.6)
\qbezier(225.9,116.6)(221.6,117.6)(217.4,118.5)
\qbezier(217.4,118.5)(213.1,119.1)(208.7,119.6)
\qbezier(208.7,119.6)(204.4,119.9)(200.0,120.0)
\qbezier(200.0,120.0)(195.6,119.9)(191.3,119.6)
\qbezier(191.3,119.6)(186.9,119.1)(182.6,118.5)
\qbezier(182.6,118.5)(178.4,117.6)(174.1,116.6)
\qbezier(174.1,116.6)(169.9,115.4)(165.8,114.0)
\qbezier(165.8,114.0)(161.7,112.4)(157.7,110.6)
\qbezier(157.7,110.6)(153.8,108.7)(150.0,106.6)
\qbezier(150.0,106.6)(146.3,104.3)(142.6,101.9)
\qbezier(142.6,101.9)(139.1,99.3)(135.7,96.6)
\qbezier(135.7,96.6)(132.4,93.7)(129.3,90.7)
\qbezier(129.3,90.7)(126.3,87.6)(123.4,84.3)
\qbezier(123.4,84.3)(120.7,80.9)(118.1,77.4)
\qbezier(118.1,77.4)(115.7,73.7)(113.4,70.0)
\qbezier(113.4,70.0)(111.3,66.2)(109.4,62.3)
\qbezier(109.4,62.3)(107.6,58.3)(106.0,54.2)
\qbezier(106.0,54.2)(104.6,50.1)(103.4,45.9)
\qbezier(103.4,45.9)(102.4,41.6)(101.5,37.4)
\qbezier(101.5,37.4)(100.9,33.1)(100.4,28.7)
\qbezier(100.4,28.7)(100.1,24.4)(100.0,20.0)

\qbezier(198.5,21)(200,22.5)(201.5,24)
\qbezier(198.5,24)(200,22.5)(201.5,21)
\qbezier(198.5,41)(200,42.5)(201.5,44)
\qbezier(198.5,44)(200,42.5)(201.5,41)
\qbezier(198.5,46)(200,47.5)(201.5,49)
\qbezier(198.5,49)(200,47.5)(201.5,46)
\qbezier(198.5,66)(200,67.5)(201.5,69)
\qbezier(198.5,69)(200,67.5)(201.5,66)
\qbezier(198.5,71)(200,72.5)(201.5,74)
\qbezier(198.5,74)(200,72.5)(201.5,71)
\qbezier(198.5,91)(200,92.5)(201.5,94)
\qbezier(198.5,94)(200,92.5)(201.5,91)
\qbezier(198.5,96)(200,97.5)(201.5,99)
\qbezier(198.5,99)(200,97.5)(201.5,96)

\put(290,130){\line(0,1){10}}
\put(290,130){\line(1,0){10}}
\put(292,132){$k$}
\end{picture}
\label{fig:2}

\para
The end result for the two terms is
\begin{eqnarray*}
\int_{-\infty}^{\infty} dk \ e^{ikr}\,\frac{\sinh 2kz^{\star} }{\cosh 2kz^{\star} - \cos \theta} & = & \frac{i \pi}{z^{\star} } \sum_{n=0}^{\infty} \left( e^{- \tfrac{r}{z^{\star}}\left(n \pi  +\tfrac{\vert \theta \vert }{2} \right) } + e^{ - \tfrac{r}{z^{\star}}\left((n + 1)\pi - \tfrac{\vert \theta \vert }{2} \right) }\right) \\
\int_{-\infty}^{\infty} dk\ e^{ikr} \,\frac{\sin\theta }{\cosh 2kz^{\star} - \cos \theta} & = & \text{ sign}(\theta)\frac{\pi }{z^{\star} } \sum_{n=0}^{\infty} \left( e^{- \tfrac{r}{z^{\star}}\left(n \pi + \tfrac{\vert \theta \vert }{2} \right)} - e^{ - \tfrac{r}{z^{\star}}\left((n + 1)\pi - \tfrac{\vert \theta \vert }{2} \right)} \right)
 \end{eqnarray*}
Summing these two series gives us the Euclidean correlator, depending explicitly on the IR boundary condition labelled by $\theta\in (-\pi,\pi]$. 
\begin{equation}
\langle \bar{\Psi}_{\beta} (\vec{0}) \Psi_{\alpha} (\vec{r}) \rangle =  -\frac{1}{4 \pi z^{\star}} \left( \vec{\sigma}_{\alpha \beta} \cdot \vec{\hat{r}} \,\frac{\partial}{\partial r} \frac{1}{r}\frac{\cosh \left( \tfrac{(\pi- \vert \theta \vert )r}{2z^{\star}} \right)}{\sinh \left( \frac{\pi r}{2z^{\star}} \right)}  + \text{ sign} (\theta)\delta_{\alpha \beta} \frac{1}{r} \frac{\partial}{\partial r}  \frac{\sinh \left( \tfrac{(\pi- \vert \theta \vert )r}{2z^{\star}} \right)}{\sinh \left( \frac{\pi r}{2z^{\star}} \right)} \right) \label{threedim}
\end{equation} 
For $\theta\neq 0$, the long-distance behaviour suffers an exponential fall-off, as befits a gapped theory. However, at $\theta=0$, the two-point function reverts to algebraic behaviour. In contrast, at short distances we  recover the conformal result,
\be
\langle \bar{\Psi}_{\beta} (\vec{0})\, \Psi_{\alpha} (\vec{r}) \rangle\ \longrightarrow\ \frac{ \vec{\sigma_{\alpha \beta}} . \hat{\vec{r}} }{\pi^{2} r^{3}} \left( 1 + {\cal O} \left( \frac{r}{z^{\star}}\right) ^{3} \right)\label{shorty}\ee

\subsubsection*{The Boundary Condensate}

Our primary interest in the following section will be  in the boundary condensate, $\langle\bar{\Psi}\Psi\rangle$, arising from taking the trace over both spinor and position indices in \eqn{corr}.  For the case of a massless fermion, it is simple to compute the condensate  using \eqn{threedim}. The first term $\sim \vec{\sigma}_{\alpha\beta}$ is killed by the trace over spinor indices. The  Laurent expansion of the trace part $\sim \delta_{\alpha \beta}$ of the two-point correlator contains even powers of $r$ and the lowest order term is of order ${\cal O}(r^{0})$. This lowest term survives in the $r \to 0$ limit and affords $\bar{\Psi} \Psi$ a vacuum expectation value. 
\begin{equation}
\langle \bar{\Psi} \Psi \rangle = \frac{\theta (\pi- \vert \theta \vert )( 2\pi-\vert \theta \vert)}{24 \pi^{2} z^{\star 3}} \label{threedimcond}
\end{equation}
The condensate vanishes when $\theta=0$ and $\theta=\pi$. Recall that these are the two special theories which preserve parity. A non-zero expectation value for $\langle \bar{\Psi}\Psi\rangle$ would  spontaneously break parity; we see that this doesn't happen.

\para
In contrast, when $\theta\neq 0,\pi$, there is no symmetry protection and the condensate is non-zero. This reflects the fact that parity is explicitly broken in such theories. 

%One can follow the condensate as $\theta \to 0$ or $\pi$ pictorially. The simple poles in Figure 2 come together to form double poles with zero residue. This is the reason for the vanishing condensate. 

\subsubsection*{A Comment on Fermions in AdS${}_2$}

In AdS${}_4$, one can follow the condensate as $\theta \to 0$ or $\pi$ pictorially. The simple poles in Figure 2 come together to form double poles with zero residue. This is the reason for the vanishing condensate. 

\para
However,  when $\theta \to 0$, there is a near-survivor: the un-partnered residue at $i \theta / 2z^{\star}$ that is becoming massless in the $\theta \to 0$ limit. But even this residue falls at the last hurdle -- it is wiped out by the $\partial_{r}$ derivative. The $\partial_{r}$ derivative is the result of  the $d=2+1$-dimensional angular integrals. Had we done the calculation in AdS$_{2}$, the $i \theta / 2z^{\star}$ residue, and hence the condensate -- now of the form $\bar{\psi}\Gamma^3\psi$ for a two-component Dirac spinor $\psi$ --- survives. Of course, in AdS${}_2$, there is no parity symmetry of the boundary theory since there is nothing to reflect. However, in this case $\bar{\psi} \Gamma^3 \psi$ is odd under charge conjugation and, at $\theta=0$, this discrete symmetry is spontaneously broken.

\para
In the following section, we will turn on a magnetic field in AdS${}_4$ and realise an effective AdS$_{2}$ geometry via formation of Landau levels. We will see that similar comments apply, this time resulting in the spontaneous breaking of CP.

\section{Magnetic catalysis}
\label{sec:4}

 In this section, we flood AdS$_{4}$ with a magnetic field. The magnetic flux spouts out from the IR hard wall and, on top of the Kaluza-Klein spectrum, a new kind of discretisation emerges in the form of Landau levels: the magnetic field slices through the dynamical phase space, reducing the degrees of freedom to layers upon layers of what is effectively AdS$_{2}$.  
  
 \para
 The dynamics of fermions in a magnetic hard wall were recently explored in \cite{magcat}, where a new form of magnetic catalysis was described. For massless bulk fermions, it was shown that, when the  IR boundary condition is given by $\theta=0$,  a bulk condensate $\langle \bar{\psi}\Gamma^5\psi\rangle$ forms. This spontaneously breaks CP symmetry\footnote{The magnetic field itself, $B = \partial_{x}A_{y} - \partial_{y}A_{x}$, is odd under parity which flips just a single spatial direction. It is also under under $C$, but preserves CP.}. In the boundary theory, this was interpreted as the formation of a $\langle \bar{\Psi}\Psi\rangle$ condensate which again spontaneously breaks CP. The bulk magnetic catalysis was therefore mirrored in a strongly interacting version of boundary magnetic catalysis. 
  
 \para
 As described in the introduction, the discussion in \cite{magcat} left two unanswered questions. Firstly, the bulk analysis relied crucially on massless fermions and, while it was conjectured that  a similar phenomenon occurs for massive fermions, no proof was given. Secondly, only heuristic arguments were given relating the bulk and boundary phenomena. 
 
 \para
The purpose of this section is to rectify both of these omissions. Using the technology developed in Section \ref{sec:2}, suitably modified to include the presence of the magnetic field, we directly compute the two-point function in the boundary theory, $\langle\bar{\Psi}(\vec{r})\,\Psi(\vec{0})\rangle$, and hence the boundary condensate.

\subsection{Landau levels in AdS$_{4}$}
\label{sub:4.1}

 One of the advantages of working with a magnetic hard wall is that one can consistently neglect the backreaction of the magnetic field on the geometry. In the full AdS geometry, this is not possible; in the deep IR the magnetic field is squeezed until it necessarily backreacts strongly on the geometry, ultimately forming a magnetic \RN black hole. Solutions to the Dirac equation in this background were studied in \cite{RN1,RN2}. However, with the hard wall in place, the IR geometry can be cut-off before the backreaction gets strong. (Of course, one is still left with the problem of describing a natural mechanism for the wall to emit a magnetic field; we will deal with this issue in Section \ref{monwallsec}).
 
 \para
 In this subsection, we start by describing the Landau levels in the full AdS${}_4$ geometry; in Section  \ref{sub:4.2} we see the effect of introducing IR boundary conditions of a hard wall.

\para 
We  apply a constant magnetic field  AdS${}_4$ in Landau gauge: $A_{y} = Bx$ with $B > 0$. We will use two commuting sets of projection operators, $P^{\pm} = \frac{1}{2}(1 \pm \Gamma^{z})$ and $Q_{(\pm)} = \frac{1}{2} (1 \pm i \Gamma^{x} \Gamma^{y})$. Their simultaneous eigenstates $\psi_{(\pm)}^{\pm}$ serve as a basis-independent decomposition of the spinor into its four components. $\psi_{(\pm)}^-$ are sources; $\psi_{(\pm)}^+$ are responses; $\psi_{(+)}^{\pm}$ are spin-up states; $\psi_{(-)}^{\pm}$ are spin-down states.

\paragraph{} The AdS$_4$ Dirac equation is minimally coupled to the magnetic field. 
\begin{equation}
z\Gamma^{a}\partial_{a} \psi - \tfrac{3}{2} \Gamma^{z} \psi - i z B x \Gamma^{y} \psi - m \psi = 0
\end{equation} 
The general solution is 
 \begin{eqnarray}
\psi(z,x,y,t) & = & \int \frac{dkd\omega}{(2\pi)^{2}} \sum_{n=0}^{\infty} e^{-i\omega t - iky} z^{2} \label{landaulev} \\
&&\left\{  Y_{m + \frac{1}{2}}(q_{n}(\omega)z) \left(a_{n(+)}^{-}(k,\omega)X_{n-1}(x,k)+a_{n(-)}^{-}(k,\omega)X_{n}(x,k)\right)\right.  \notag \\
&& \quad +Y_{m - \frac{1}{2}}(q_{n}(\omega)z) \left(a_{n(+)}^{+}(k,\omega)X_{n-1}(x,k)+a_{n(-)}^{+}(k,\omega)X_{n}(x,k)\right) \notag \\
&& \quad +J_{m + \frac{1}{2}}(q_{n}(\omega)z) \left(b_{n(+)}^{-}(k,\omega)X_{n-1}(x,k)+b_{n(-)}^{-}(k,\omega)X_{n}(x,k)\right) \notag \\
&& \quad \left. +J_{m - \frac{1}{2}}(q_{n}(\omega)z) \left(b_{n(+)}^{+}(k,\omega)X_{n-1}(x,k)+b_{n(-)}^{+}(k,\omega)X_{n}(x,k)\right) \right\} \nn\end{eqnarray}
Here, $q_{n}(\omega) \equiv \sqrt{\omega^{2} - 2Bn}$, and 
\begin{equation*}
X_{n}(x,k) \equiv \left( \frac{\sqrt{B}}{2^{n} n! \sqrt{\pi}} \right)^{\frac{1}{2}} \exp \left( - \frac{B}{2} \left(x+ \frac{k}{B} \right) ^{2} \right) H_{n} \left( \sqrt{B}\left(x + \frac{k}{B} \right) \right), \end{equation*}
 are the normalised wavefunctions of the simple harmonic oscillator wavefunctions. $X_{-1}=0$ is an empty placeholder; similarly, we define $a_{0(+)}^\pm(k, \omega), b_{0(+)}^\pm(k, \omega) \equiv 0$. The other plus and minus spinor components are related by the condition
\begin{eqnarray*}
(a_{n(-)}^{+}(k,\omega) + a_{n(+)}^{+}(k,\omega)) & = & - \left( \frac{i\omega}{q_{n}(\omega)} \Gamma^{t} + \frac{i \sqrt{2Bn}}{q_{n}(\omega)} \Gamma^{y} \right) (a_{n(-)}^{-}(k,\omega) + a_{n(+)}^{-}(k,\omega)) \\
(b_{n(-)}^{+}(k,\omega) + b_{n(+)}^{+}(k,\omega)) & = & - \left( \frac{i\omega}{q_{n}(\omega)} \Gamma^{t} + \frac{i \sqrt{2Bn}}{q_{n}(\omega)} \Gamma^{y} \right) (b_{n(-)}^{-}(k,\omega) + b_{n(+)}^{-}(k,\omega)).
\end{eqnarray*} 
%
%
% Verifying this solution requires the properties $\Gamma^z \psi^\pm = \pm \psi^\pm$ and $\Gamma^x \psi_{(\pm)} = \pm i \Gamma^y \psi_{(\pm)}$. In addition to the Bessel function identities of Section~\ref{sub:2.2}, we also need the Hermite polynomial identities $H_n'(u) = 2n H_{n-1}(u)$ and $H_{n-1}'(u) = 2uH_{n-1}(u) - H_n(u)$.
%
%\paragraph{} The solution \eqref{landaulev} is valid for $B > 0$. For $B < 0$, the spin-up states $a^\pm_{n(+)}(k, \omega)$, $b^\pm_{n(+)}(k, \omega)$ swap roles with the spin-down states $a^\pm_{n(-)}(k,\omega)$, $b^\pm_{n(-)}(k,\omega)$. Factors of $\sqrt{2Bn}$ become $-\sqrt{2Bn}$, and occurences of $(x + k/B)$ in the definition of $X_n(x,k)$ turn into $(x - k/B)$. 

\para
The AdS solution resembles its Minkowski space cousin. The magnetic field has reduced the number of true, continuous dynamical dimensions by two. In place of a momentum conjugate to $x$, there is a discrete sum over Landau levels. Meanwhile, the momentum $k$, conjugate to $y$, has had its dynamical role stolen by $\sqrt{2Bn}$.

\paragraph{} The energy of the $n^{\text{th}}$ relativistic Landau level is $\sqrt{2Bn}$. The lowest, $n=0$   Landau level is special in two respects: it has half the states of the Landau levels and it has zero energy.

\para
There is a simple, intuitive explanation for both of these facts, arising from the spin coupling to the magnetic field, which raises the energy of the spin-up  state and lowers the energy of each spin-down state. The $n^{\text{th}}$ relativistic Landau level is the confluence where the raised $(n-1)^{\text{th}}$ spin-up state meets the lowered $n^{\text{th}}$ spin-down state. The lowest relativistic Landau level contains the $0^{\text{th}}$ spin-down state but it does not contain a spin-up state at all. Furthermore, the reduction in energy due to the spin coupling exactly cancels the usual ground state energy of the non-relativistic Landau level, resulting in a ground state with vanishing energy.

\subsection{Reintroducing the IR Hard Wall}
\label{sub:4.2}

We now put the hard wall back, cutting off the infra-red of the geometry at $z=\zstar$. We again work with the one-parameter family of boundary conditions \eqn{sir} labelled by $\theta\in (0,\pi]$. The computation of the boundary two-point function follows that given in \ref{sub:2.3}. We relegate details to Appendix \ref{magwallapp}, here quoting only the answer.

\para
The spectrum of the theory, defined by \eqn{spec} in the absence of a magnetic field, now becomes
\begin{equation}
\frac{J_{m+\frac{1}{2}}(z^{\star} \sqrt{\omega^{2}-2Bn} )}{J_{m-\frac{1}{2}}(z^{\star} \sqrt{\omega^{2}-2Bn})} = \pm \tan \tfrac{\theta}{2}
\end{equation} 
When $\theta=0$, there is once again a zero energy mode ($\omega=0$), now lying in the lowest $n=0$ Landau level.  In fact, more precisely, there is a flat band of such modes, labelled by $k$. As we will see shortly, this lowest Kaluza-Klein mode within the lowest Landau level will be the key player in our story.

\para
For general $\theta$, the two-point correlation function is most is simply written when $(\vec{x}-\vec{y})$ has only temporal separation, $t$. (See Appendix \ref{magwallapp} for the more general form). Since we are ultimately interested in the condensate $\langle\bar{\Psi}\Psi\rangle$, we further take the spinor trace of the two-point function. This is given by
\begin{multline}
\langle\bar{\Psi}(0,0,0) \Psi(0,0,t) \rangle =  -\frac{i B}{2\pi} \int_{-\infty}^{\infty} \frac{d\omega}{2\pi} e^{- i \omega t} \left( \left( \frac{\vert \omega \vert}{2} \right)^{2m} \left( - \text{sign}(\omega) N_{1}(\vert \omega \vert) + N_{2}(\vert \omega \vert)\right)  \right. \\
+ \left. 2 \sum_{n=1}^{\infty} \left( \frac{q_{n}(\omega)}{2}\right)^{2m} N_{2}(q_{n}(\omega)) \right) \label{mag}
\end{multline}
with the functions $N_1$ and $N_2$ defined in \eqn{coeffs2}.

\para
The top line of \eqref{mag} comes from the lowest Landau level. It is the two-point function $\langle \bar{\Psi}(0) \Psi(t) \rangle$ that we would have calculated in Section~\ref{sub:2.3}, had we been working in AdS$_{2}$ rather than AdS$_{4}$. 
%($N_{1}(q_n (\omega))$ and $N_{2}(q_n (\omega))$ appear side by side in lieu of a traceless gamma matrix in $0+1$ dimensions.) 
The $B / 2\pi$ factor is the degeneracy of the lowest Landau level.

\para
The  bottom line of \eqref{mag} contains contributions from higher Landau levels. They have double the occupancy of the lowest Landau level. The $N_{1}(q_n (\omega))$ terms cancel in pairs to leave only the $N_{2}(q_n (\omega))$ terms, similar to the result \eqref{threedim} for fermions in the absence of the $B$ field.

\subsection{Spontaneous CP Breaking}
\label{sub:t4.3}

We now turn to the phenomenon of magnetic catalysis, the spontaneous breaking of CP symmetry through the formation of a condensate $\langle\bar{\Psi}\Psi\rangle$. As we will see below, the breaking is due solely to the zero mode that appears in the $\theta=0$ theory.

\subsubsection*{The usual warm-up with the massless bulk fermion}

Before discussing general bulk masses $m$, we start with the more tractable case of $m=0$. The contribution of the lowest Landau is provided by the top line of \eqn{mag}. We perform a Wick rotation to Euclidean time $\tau$. The Fourier integrals are the ones we already performed in Section~\ref{sub:2.3}, albeit with no three-dimensional angular integrals. This means that we have no $\partial_r$ derivatives and contribution to the two-point function from the lowest Landau level is
\begin{eqnarray}
\langle \bar{\Psi} (0) \Psi (\tau) \rangle_{n=0} & = & \frac{B}{2\pi}\int_{-\infty}^{\infty} \frac{d \omega}{2 \pi} e^{i \omega \tau} \frac {-i \sinh 2 \omega z^{\star} + \sin \theta}{\cosh 2 \omega z^{\star} - \cos \theta} \notag \\
& = & \frac{B}{4\pi z^{\star}} \left(\frac{ \cosh \left( \tfrac{(\pi-\vert \theta \vert ) \tau}{2z^{\star}} \right)}{ \sinh \left( \frac{\pi \tau}{2z^{\star}} \right)} + \text{sign}(\theta) \frac{\sinh \left( \tfrac{(\pi- \vert \theta \vert ) \tau }{2z^{\star}} \right)}{\sinh \left( \frac{\pi \tau}{2z^{\star}} \right)} \right) \label{onedim}
\end{eqnarray}
The first term in \eqref{onedim} is odd in $\tau$ and, moreover, diverges as $\tau\rightarrow 0$. In the absence of the magnetic field, the divergent term occurred only in the traceless part of the correlator \eqn{threedim} and disappeared upon taking the trace over spinor indices. 
The analogous thing to do here is to temporarily Wick rotate back to the Lorentzian signature and keep only the \emph{real} part of $\bar{\Psi}(0) \Psi (t)$. ($\bar{\Psi} \Psi$ is a real bilinear in the Lorentzian signature.)

\paragraph{} That leaves the second term in \eqref{onedim}, which is even in $\tau$. This term has no divergence as $\tau \to 0$, and it does acquires a finite condensate.
\begin{equation}
\langle \bar{\Psi} \Psi \rangle = \text{sign}(\theta)\,\frac{(\pi - \vert \theta \vert ) }{4 \pi^{2} z^{\star}}\,B\label{cond}
\end{equation}
For $\theta\neq 0,\pi$, the boundary conditions break CP and the presence of the condensate is no surprise. Indeed, we have already seen that such a condensate is also present for $\theta \neq 0,\pi$ in the absence of a $B$ field. (The higher Landau levels contributing terms which do not vanish in the limit $B\rightarrow 0$).

\para
The surprise occurs at in the limit $\theta\rightarrow 0$. Here the theory is CP invariant, but the condensate does not vanish. 
CP is spontaneously broken.
\begin{equation}
\lim_{\theta \rightarrow 0^\pm} \,\langle \bar{\Psi} \Psi \rangle = \frac{B}{4 \pi z^{\star}}\, \text{sign}(\theta)\label{lovely}
\end{equation}
The magnitude of the condensate is proportional to both the spectral mass separation, $1/\zstar$, and the magnetic field strength.

\para
It is not hard to see where the condensate gets its contribution; it comes from the lone residue  $i \theta / 2 z^{\star}$ pole, the lowest excitation that becomes massless as $\theta \to 0$. As we explained in Section \ref{subway}, in the absence of the magnetic field the residue of this pole succumbed  to a $\partial_r$ derivative arising from angular integration. The magnetic field effectively reduces the dynamics on the boundary to quantum mechanics. With only one dynamical boundary dimension, there are no angular integrals and no $\partial_{r}$ derivative, and the $i \theta / 2 z^{\star}$ residue survives. We can also look at the contribution from the other poles, representing the higher Kaluza-Klein levels (still within the lowest Landau level). As $\theta\rightarrow 0$, the poles of the higher excitations come together to form double poles with zero residue. This ensures that only the lowest Kaluza-Klein excitation  contributes to the condensate.

\para
So far, we have only considered the contribution from the lowest Landau level. There is the question of whether the higher Landau levels also contribute to the condensate in the $\theta \to 0$ limit. The answer is that they do not. It will be convenient to defer the proof until after we have treated the case where the bulk fermion can take general mass.

\para
Finally, we should mention that although we were ultimately interested in the $\theta=0$ theory, it was necessary to construct the whole circle of IR boundary conditions.  Had we set $\theta = 0$ in the beginning, we would never have seen the $\sin \theta$ term in \eqref{onedim}, and we would have missed the CP-breaking condensate altogether. This is a familiar story in mean field phase transitions; it corresponds the sitting in an unstable vacuum, perched precariously on at the maximum of the potential. Turning on a small $\theta$, we slid down the hillside to the valley, and sending $\theta \to 0$, we recovered the theory we were interested in, nested in the safety of the true vacuum.

\subsubsection*{Generalising to arbitrary masses of the bulk fermion}

We would now like to compute the condensate for a bulk fermion of arbitrary mass $m>-\ft12$, corresponding in the boundary theory to a spinor operator with dimension greater than the free value. Unfortunately in this case, the Fourier transform is no longer tractable. Nonetheless,  we have a cunning trick up our sleeves that allows us to extract the condensate in the  $\theta \rightarrow 0$ limit. 

\para
The condensate comes from the $N_{2}$ term in the top line of \eqn{onedim}. From  \eqref{coeffs2} we see that $N_2$ is proportional to $\sin\theta$; it seems at first glance like it vanishes for $\theta = 0$. But appearances are deceptive. There is an opposing effect,  a divergence at $\omega = 0$ coming from the denominator.  Of course, this divergence can be traced back to the massless mode in the lowest Landau level: as $\theta$ approaches $0$, the pole of the lowest mode bears down onto the real axis and digs up a spike in the middle of our integration path.
%\footnote{We are not using the residue theorem here. We are merely integrating along the real axis; there is no attempt to close the contour in a semi-circle. Also remember that we performed a Wick rotation, and the non-zero spectral poles are lined up along the imaginary axis, out of harm's way.}

\para
This means that the only part of the integral  \eqn{onedim} that contributes to the $\theta=0$ condensate arises near $\omega=0$. 
We may therefore replace the integrand with the first-order expansions in both $\theta$ and $\omega$. After a Wick rotation, the integral evaluates to 
\begin{eqnarray}
\lim_{\theta \rightarrow 0} \ \langle \bar{\Psi} \Psi \rangle & = &  \lim_{\theta, \tau \to 0} \ \frac{B}{2\pi}\int_{-\infty}^{\infty} \frac{d \omega}{2 \pi} \left( \frac{\vert \omega \vert}{2} \right)^{2m} \frac{e^{i \omega \tau}}{|\omega|\zstar} \,\frac{1}{\Gamma(m+\tfrac{1}{2})^{2} } \notag \\ & & \qquad \qquad \quad \times \frac {\sin \theta }{ \cos^{2} (\tfrac{\theta}{2}) I_{m+\frac{1}{2}}^{2}(\vert \omega \vert z^{\star}) + \sin^{2} (\tfrac{\theta}{2}) I_{m-\frac{1}{2}}^{2}(\vert \omega \vert z^{\star}) } \notag
\\ & = & \lim_{\theta \rightarrow 0} \ \frac{B}{4 \pi z^{\star 2m}} \int_{-\infty}^{\infty}\ \frac{d \omega}{2 \pi} \frac{\theta}{\left( \frac{\omega z^{\star}}{2m+1} \right)^{2}+ \left( \frac{\theta}{2} \right) ^{2} } \label{inttodo} \ee
Performing this final integral, we get the expression for the condensate for general bulk mass $m$, 
\be \lim_{\theta \rightarrow 0^\pm} \,\langle \bar{\Psi} \Psi \rangle=
(2m+1)\, \frac{B}{4 \pi z^{\star 2m+1}} \text{sign} (\theta). \label{gencond}
\ee 

\subsubsection*{Higher Landau levels}
At this stage, it is easy to see that higher Landau levels make no contribution to $\langle \bar{\Psi} \Psi \rangle$. The $2Bn$ term in the Wick-rotated $q_n (\omega) \equiv \sqrt{\omega^2 + 2Bn}$ diffuses out the potential divergence at $\omega = 0$.

\subsubsection*{Comparison to Weak Coupling}

The phenomenon of magnetic catalysis was observed long ago in weakly coupled field theory in $d=2+1$ dimensions with a massless fermion $\Psi$ \cite{magcat0,magcat1,magcat2,magcat3}. In that case, in the presence of a magnetic field, the condensate remains as the mass, $M$, of the fermion is taken to zero,
\be \lim_{M \to 0^\pm} \langle \bar{\Psi} \Psi \rangle = \frac{B}{ 4\pi}\,\sign(M) . \nn\ee
Our result \eqn{gencond} is the same phenomenon in a strongly coupled framework. At weak coupling, the dimension of the fermion is $\Delta[\Psi]=1$. At strong coupling, the dimension of our fermion operator is $\Delta[\Psi]=\ft32 +m$. The condensate remains proportional to $B$, with the shortfall in dimensions made up by the appropriate power of the spectral spltting, $1/z^\star$.

\para
The results of this section suggest that the phenomenon of holographic magnetic catalysis in a $d=2+1$ dimensional boundary  appears to be a robust phenomenon. Just as in the weakly coupled case, a magnetic field is needed to reduce the effective dynamics to a zero-energy Landau level. The new ingredient that is needed is the appearance of a discrete, but gapless, mass spectrum for the fermion.

%\paragraph{Remark: When is it fair to cheat?}
%
%\paragraph{} We concocted a non-dynamical metric and gauge field. We tracked our fermion as it darted across an inert canvas, bouncing back and forth between perfect, unyielding mirrors. What is missing is the back-reaction of the fermion field onto the metric and the gauge field, and the interplay between the three fields. In other words, we created Landau levels, but we neglected the deformation of the Landau level states as they fill up. What we have done is acceptable at zero chemical potential, but it could potentially compromise the validity of the results at non-zero chemical potential.
%
%\paragraph{} There is no need to despair in the face of this adversity. The hard wall calculation has already taught us the key elements in the CP-breaking mechanism: a three-dimensional boundary, a zero mode and a relativistic dispersion. The back-reaction may change the exact radial profile of the fermion field, but it would be peculiarly unnatural if the back-reaction destroys any of these key elements. Even if the back-reaction modifies the exact expressions for the correlators, the essential CP-breaking mechanism should stay intact.

\section{Condensates in Bulk and Boundary}
\label{sec:3}

In this section, we change our focus somewhat. Rather than computing boundary condensates $\langle \bar{\Psi}\Psi\rangle$, we will instead focus on condensate of bulk fermion fields, $\langle \bar{\psi}\psi\rangle$ and $\langle\bar{\psi}\Gamma^5\psi\rangle$.  The question we wish to ask is: what is the signature of the bulk condensates from the perspective of the boundary theory? 

\para
In Section \ref{sec:4} we have seen that, under certain circumstances, a boundary $\langle \bar{\Psi}\Psi\rangle$ condensate forms in the boundary theory. In previous work \cite{magcat}, it was shown that, under the {\it same} circumstances, a bulk fermionic $\langle\bar{\psi}\Gamma^5\psi\rangle$ forms with a particular profile in the radial direction. For a massless bulk fermion, 
\be \lim_{\theta\rightarrow 0}\, \langle\bar{\psi}\Gamma^5\psi\rangle  =  \frac{B}{4\pi\zstar}\,z^3\,\sign(\theta)\nn\ee
It was suggested in \cite{magcat} that the bulk and boundary composite operators are dual,
\be \bar{\psi}\Gamma^5\psi \ \longleftrightarrow \ \bar{\Psi}\Psi\label{ilikethis}\ee
This is an extension of the usual AdS/CFT dictionary, which relates single trace operators in the boundary to single fields in the bulk.
In this section, we will provide further, strong evidence for the above relationship.

\para
Of course, there is already a well-established method for introducing the {\it source} of double trace operators involving mixed boundary conditions for the bulk fields \cite{witten,berkooz}. We will see that our proposal below is fully consistent with the results of 
 \cite{witten,berkooz}. Moreover, it allows us to identify the {\it response} $\langle\bar{\Psi}\Psi\rangle$ as the appropriate fall-off of the  composite bulk field $\bar{\psi}\Gamma^5\psi$ close to the boundary.

\para
Our analysis in this section will focus on the hard wall in the absence of a magnetic field. As we showed in Section 2, the one-parameter family of boundary conditions \eqn{sir} gives rise to a boundary condensate $\langle\bar{\Psi}\Psi\rangle$ whenever $\theta\neq 0,\pi$. We will see how to extract this from the bulk field  $\bar{\psi}\Gamma^5\psi$. Moreover, we will provide an interpretation for the other bulk condensate $\langle \bar{\psi}\psi\rangle$.

%We will show that the double-trace boundary operators $\bar{\Psi} \Psi$ and $\tfrac{1}{2} \bar{\Psi} (\gamma^i \overrightarrow{\partial_i} - \overleftarrow{\partial_i} \gamma^i) \Psi$ whose VEVs we painstakingly evaluated in Section \ref{sub:2.3} are each dual to two-particle fields in the bulk: $i\bar{\psi} \Gamma^{5} \psi$ and $\bar{\psi} \psi$. The VEVs are related, up to a conformal factor. We wonder whether we are catching a first glimpse of uncharted territory -- whether these results are a part of a wider scheme in the AdS/CFT dictionary.

\subsection{Bulk to Bulk Green's Functions}

We start by reviewing the bulk-to-bulk propagator, $G(z,\vec{x};w,\vec{y})$, for spinor fields in AdS${}_4$. This Green's function solves
\begin{equation*}
\left( e_{a}^{\mu} \Gamma^{a} \stackrel{\rightarrow}{{D}}_{(z,\vec{x})\, \mu}-m \right) G  =  G\left( - \stackrel{\leftarrow}{D}_{(w,\vec{y}) \,\mu}e_{a}^{\mu} \Gamma^{a} -m \right) =  \frac{1}{\sqrt{-g}}\delta(z-w)\delta^{3}(\vec{x}-\vec{y}),
\end{equation*}
subject to the UV boundary condition
\begin{equation*}
G \vert_{z=0} = G \vert_{w=0} =0
\end{equation*}
together with an IR boundary condition inherited from \eqn{wallbc}, 
\begin{equation*}
\left(\cos \tfrac{\theta}{2} P^{-} + i \sin \tfrac{\theta}{2} \Gamma^{5} P^{+} \right) G \vert_{z= z^{\star}} = G \vert_{w= z^{\star}} \left(\cos \tfrac{\theta}{2} P^{+} + i \sin \tfrac{\theta}{2} P^{-} \Gamma^{5}  \right) = 0.
\end{equation*}
Following \cite{kawano}, we obtain the following propagator.
\begin{multline} 
G(z,\vec{x},w,\vec{y})=\int \frac{d^{3}k}{(2\pi)^{3}} \left( - \frac{\pi k}{2} \right) e^{-i \vec{k}\cdot(\vec{x}-\vec{y})} \left( \theta(z-w) \phi_{IR}(z,\vec{k})\, P^{-} i \vec{\Gamma}\cdot\hat{\vec{k}} \,\bar{\phi}_{UV}(w,\vec{k}) \right. \\
\left. +\theta(w-z) \phi_{UV}(z,\vec{k})\, P^{-}  i \vec{\Gamma}\cdot \hat{\vec{k}}\, \bar{\phi}_{IR}(w,\vec{k}) \right),\nn
\end{multline} 
where $\theta(x)$ is the Heaviside step function and 
\begin{eqnarray*}
\phi_{UV}(z, \vec{k}) & \equiv & \bar\phi_{UV}(z, \vec{k}) \equiv z^{2} \left( J_{m+\frac{1}{2}}(kz) - i \vec{\Gamma} . \hat{\vec{k}} J_{m-\frac{1}{2}}(kz)\right) \\
\phi_{IR}(z, \vec{k}) & \equiv & z^{2} \left[ \left( Y_{m+\frac{1}{2}}(kz) - i \vec{\Gamma} . \hat{\vec{k}} Y_{m-\frac{1}{2}}(kz) \right) \right. \\ & & \qquad \qquad \qquad \qquad \left.+ \left( J_{m+\frac{1}{2}}(kz) - i \vec{\Gamma} . \hat{\vec{k}} J_{m-\frac{1}{2}}(kz) \right) (M_1(k)+\Gamma^5\vec{\Gamma}.\hat{\vec{k}}M_2 (k))\right] \\
\bar\phi_{IR}(z, \vec{k}) & \equiv & z^{2} \left[ \left( Y_{m+\frac{1}{2}}(kz) - i \vec{\Gamma} . \hat{\vec{k}} Y_{m-\frac{1}{2}}(kz) \right) \right. \\ & & \qquad \qquad \qquad \qquad \left.+ (M_1(k)-\Gamma^5\vec{\Gamma}.\hat{\vec{k}}M_2 (k)) \left( J_{m+\frac{1}{2}}(kz) - i \vec{\Gamma} . \hat{\vec{k}} J_{m-\frac{1}{2}}(kz) \right)\right]
\end{eqnarray*}
Our propagator  satisfies the hermiticity condition $\Gamma^0 G(z,\vec{x},w,\vec{y})^\dagger \Gamma^0= -G(w,\vec{y},z,\vec{x})$.

\subsection{Identifying $ \bar{\psi} \Gamma^5 \psi$}

We start by computing the condensate $\langle i \bar{\psi} \Gamma^5 \psi \rangle$ in the hard wall background as a function of the IR boundary condition $\theta$. The calculation is simplest for the massless bulk fermion since the Fourier transforms reduce to those already computed in Section \ref{subway}. After a Wick rotation, we find\footnote{$\langle i \bar{\psi} (z, \vec{y}) \Gamma^5 \psi (z, \vec{x}) \rangle$ has no divergence as $\vec{x} \to \vec{y}$, but the full $\langle \bar{\psi}_{\beta} (z, \vec{y}) \psi_{\alpha} (z, \vec{x}) \rangle$ does. There is a standard procedure for renormalizing divergences that occur in the expectation values of composite operators in curved spacetime; the procedure is described in \cite{birrell}. One may ask whether the renormalization procedure could subtract a finite piece away from our expression for $\langle i \bar\psi \Gamma^5 \psi \rangle$. We can check  that this does not occur by examining the gamma matrix structure of the expression that is to be subtracted, which is equation (6.92) in \cite{birrell}.}
 \begin{eqnarray}
\langle i\bar{\psi}(z,\vec{x})\Gamma^{5} \psi (z, \vec{x}) \rangle  = - \text{ tr} \left(G(z,\vec{x}, z, \vec{x}) \Gamma^{5} \right)
%& = & \lim_{r \to 0}\frac{z^{3}}{2\pi^2 r} \frac{\partial}{\partial r} \Re \int_{-\infty}^{\infty} dk e^{ikr} \frac{\cosh 2kz \sin\theta}{\cosh 2kz^{\star} - \cos \theta} \notag \\
%& = & \text{sign}(\theta) \frac{z^3}{4\pi z^\star} \lim_{r \to 0} \frac{1}{r} \frac{\partial}{\partial r} \left( \frac{\sinh \left(\frac{(\pi - \vert \theta \vert ) (r - 2iz) }{2z^\star} \right)}{\sinh \left( \frac{\pi (r - 2iz)}{2z^\star} \right)} + \frac{\sinh \left(\frac{(\pi - \vert \theta \vert ) (r + 2iz) }{2z^\star} \right)}{\sinh \left( \frac{\pi (r + 2iz)}{2z^\star} \right)} \right) \notag \\
 =  -\text{sign}(\theta) \,\frac{z^3}{8 \pi z^{\star}} \frac{\partial^2}{\partial z^2} \frac{\sin \left(\tfrac{(\pi - \vert \theta \vert) z}{z^{\star}}\right)}{\sin \left(\tfrac{\pi z}{z^{\star}}\right)}\ \  \label{bulkcond}
\end{eqnarray}
Expanding the bulk condensate near the boundary $z=0$, we find its leading behaviour 
\begin{equation*}
\langle i \bar{\psi} \Gamma^5 \psi \rangle = - \frac{ \theta (\pi - \vert \theta \vert ) (2 \pi - \vert \theta \vert)}{24 \pi^2 z^{\star 3}}\,z^3 + O(z^5)
\end{equation*}
We see that the leading order behaviour of the bulk condensate coincides with the boundary condensate that we computed in \eqn{threedimcond}. We can write,
\be
\langle i \bar{\psi} \Gamma^{5} \psi \rangle=-\langle\bar{\Psi} \Psi \rangle\,z^{3}+\ldots   \nn
\ee
Note that both bulk and boundary condensates have the same transformation properties under the discrete symmetries: both are odd under P, even under C.

\subsubsection*{More General Bulk Masses}

For more mass $m\neq 0$, we can no longer do the Fourier transform to explicitly compute the bulk condensate. Nonetheless, we can still relate it to the integral representation of the boundary condensate \eqn{corr}. Again performing a Wick rotation, the bulk condensate takes the (finite) form
\be
\langle i \bar{\psi}(z, \vec{x}) \Gamma^5 \psi(z, \vec{x}) \rangle = 2z^4 \int \frac{d^3k}{(2\pi)^3} k \left( I_{m+\frac{1}{2}}^2(kz)+ I_{m-\frac{1}{2}}^2(kz)\right) {\cal M}_2 (k)
\ee 
where ${\cal M}_2$ can be thought of as the Euclidean version of $M_2$ appearing in \eqn{coeffs}
\begin{equation*}
{\cal M}_2 (k) \equiv -\frac{1}{2k\zstar}\frac{\sin\theta } {\cos^2\tfrac{\theta}{2} I_{m+\frac{1}{2}}^2 (kz^\star) + \sin^2 \tfrac{\theta}{2} I_{m-\frac{1}{2}}^2(kz^\star)}.
\end{equation*}
Close to the boundary of AdS, we can expand this as
\be
\langle i \bar{\psi} \Gamma^5 \psi \rangle = \int \frac{d^3k}{(2\pi)^3} \left( \frac{k}{2} \right)^{2m} \frac{2 {\cal M}_2(k)}{\Gamma (m + \tfrac{1}{2})^2 }\,z^{3+2m}+\ldots\nn
\ee
A short calculation confirms that this leading order term agrees with the boundary condensate derived from \eqn{corr}. We have established the following result.
\be
\langle i \bar{\psi} \Gamma^{5} \psi \rangle=-\langle\bar{\Psi} \Psi \rangle\,z^{3+2m}+\ldots   \label{genm5}
\ee 
The result has a conformal factor of $z^{3+2m}$, in accordance with the conformal dimension $3+2m$ of the operator $\bar{\Psi} \Psi$. Note that both operators are odd under P and even under C, so the discrete symmetries match.

\para
In Appendix \ref{hideapp}, we will show this correspondence relies at heart on nothing more than the fact that bulk-to-boundary propagators are limits of bulk-to-bulk propagators.

\subsubsection*{Comparison to a Bulk Pseudo-Scalar}

We can compare the result \eqn{genm5}  to the familiar case of a massless bulk pseudo-scalar field, $\phi$. If the scalar has mass $M^2=2m(3+2m)$, the leading and subleading fall-offs take the form,
\be \phi(\vec{x},z) = J(\vec{x})\,z^{-2m} + O(\vec{x})\,z^{3+2m} + \ldots
\nn\ee
where $J(\vec{x})$ is interpreted as the source for the dual operator, while $O(\vec{x})$ is the response. We wish to argue that the composite fermion field $i\bar{\psi}\Gamma^5\psi$ acts rather like such a pseudo-scalar, dual to the double trace operator $\bar{\Psi}\Psi$. Certainly, as we have seen, the subleading fall-off of the bulk condensate indeed captures the boundary condensate $\langle\bar{\Psi}\Psi\rangle$. What about the source? 

\para
In fact, for massless bulk fermions, the source for $\bar{\Psi}\Psi$ can indeed be shown to induce a leading, constant term in the bulk condensate $\langle i\bar{\psi}\Gamma^5\psi\rangle$. This follows from an old calculation \cite{allen} which shows that $\bar{\psi}\psi$ has a non-zero, constant expectation value in AdS${}_4$. (We will revisit this calculation in the following Section \ref{seccond}). 
It is known that a source can be turned on by changing the UV boundary conditions \cite{witten,berkooz} which, for a massless bulk fermion, is a marginal deformation in the boundary theory. The result is that the constant $\langle\bar{\psi}\psi\rangle$ condensate is rotated into a $\langle \bar{\psi}\Gamma^5\psi\rangle$ condensate \cite{magcat}. 

\para
For $m\neq 0$, a source for the double trace operator is either irrelevant (for $m>0$), destroying the UV geometry, or relevant (for $m<0$), where it institutes an RG flow \cite{allais,laia2,gautam}. It would be interesting to check in these cases if the source again affects the appropriate leading terms of the bulk condensate.

\subsection{Identifying $\bar{\psi} \psi$}
\label{seccond}

We now turn to the  bulk condensate $\langle\bar{\psi}\psi\rangle$. Once again, we would like to understand the image of this condensate in the boundary theory. Below, we will show that this bulk composite field is again dual to a boundary double trace operator,
\be \bar{\psi}\psi\ \longleftrightarrow\  \bar{\Psi}\!\!\!\stackrel{\ \, \leftrightarrow}{\delslash}\!\Psi\nn\ee
where $\bar{\Psi}\!\!\!\stackrel{\  \leftrightarrow}{\delslash}\!\Psi\equiv \ft12 \bar{\Psi}(\! \!\stackrel{\ \, \rightarrow}{\delslash}-\stackrel{\ \leftarrow}{\delslash})\Psi$. Note that both bulk and boundary operators are even under both P and C. 

\subsubsection*{A Massless Bulk Fermion}

We once again start by looking at the case of a massless bulk fermion where explicit expressions are available. 
However, in this case an old calculation, predating the AdS/CFT correspondence, shows that there is a bulk condensate $\langle\bar{\psi}\psi\rangle$ even in pure AdS without a hard wall,
\be \langle\bar{\psi}\psi\rangle = \frac{1}{4\pi}\label{4pi}\ee
This can be understood as a consequence of the chiral symmetry breaking boundary conditions which are necessarily imposed on the UV boundary of AdS. This explicit chiral symmetry breaking infects the bulk of AdS through the presence of the constant condensate\footnote{A particularly simple derivation uses the fact that the massless fermion is conformal to a fermion in flat half-space, with chiral symmetry breaking boundary conditions. (See, for example, \cite{magcat}). Once again, one may ask whether renormalization could subtract a finite piece from \eqn{4pi}. To show that this does not occur, we use the chiral symmetry of the theory of the massless fermion to rotate $\bar{\psi}\psi$ into $i\bar{\psi}\Gamma^5\psi$ \cite{magcat}. We have already checked that renormalization does not lead to a subtraction from $i\bar\psi \Gamma^5 \psi$. Although chiral rotations change the boundary conditions in the IR and the UV, the bulk subtractions depend only on local properties of the theory, not on the boundary conditions \cite{birrell}. In the renormalization procedure described in \cite{birrell}, the subtractions are the first three leading terms in the bulk Green's function, expanded as a series in geodesic distance between the two points at which the Green's function is evaluated. Since bulk subtractions are determined in this geometric way, they should transform in the expected manner under chiral rotations. We already know that there is no subtraction from $i\bar\psi \Gamma^5 \psi$, so by performing a chiral rotation, we infer that there is no subtraction from $\bar\psi \psi$ either. Thus we conclude that \eqn{4pi} is indeed the correct result.}

\para
We can also compute the condensate in the presence of the hard wall with boundary condition labelled by $\theta$. The calculation is superficially similar to those of the previous section, albeit with different UV behaviour. We find
\begin{eqnarray}
\langle \bar{\psi} (z, \vec{x}) \psi (z, \vec{x})\rangle  =  i \text{ tr} \left( G(z,\vec{x}, z, \vec{x}) \right) 
%& = & \lim_{r \to 0}\frac{ z^{3}}{2\pi^2 r} \frac{\partial}{\partial r} \Re \int_{-\infty}^{\infty} dk e^{ikr} \frac{\cosh 2k z \cos \theta- \cosh 2k(z - z^{\star})}{\cosh 2kz^{\star} - \cos \theta} \notag \\
%& = & \frac{i z^3}{4\pi z^\star} \lim_{r \to 0} \frac{1}{r} \frac{\partial}{\partial r} \left( \frac{\cosh \left(\frac{(\pi - \vert \theta \vert ) (r - 2iz) }{2z^\star} \right)}{\sinh \left( \frac{\pi (r - 2iz)}{2z^\star} \right)} - \frac{\cosh \left(\frac{(\pi - \vert \theta \vert ) (r + 2iz) }{2z^\star} \right)}{\sinh \left( \frac{\pi (r + 2iz)}{2z^\star} \right)} \right) \notag \\
= \frac{z^3}{8 \pi z^{\star}} \frac{\partial^2}{\partial z^2} \frac{\cos\left( \tfrac{(\pi - \vert \theta \vert ) z}{z^{\star}}\right)}{\sin \left(\tfrac{\pi z}{z^{\star}}\right)}\label{bouncey}
\end{eqnarray}
Near the UV boundary, the condensate has the following behaviour. 
\be
\langle \bar{\psi} \psi \rangle = \frac{1}{4 \pi^2} + \frac{15 \vert \theta \vert ^4 - 60 \vert \theta \vert ^3 \pi + 60 \vert \theta \vert ^2 \pi^2 - 8 \pi^4}{480 \pi^2 z^{\star 4}} \,z^4 + O(z^6) \label{goodf}
\ee
We would like to identify the meaning of the subleading, $z^4$ term in the boundary theory.  

\para
To do this, let's look again at the boundary two-point function \eqn{threedim}. Taking a derivative, the short-distance expansion is given by
\begin{equation*}
 \langle \bar{\Psi} (\vec{0}) \sigma^i \partial_i \Psi (\vec{r}) \rangle = -\frac{2}{\pi^2 r^4} - \frac{15 \vert \theta \vert ^4 - 60 \vert \theta \vert ^3 \pi + 60 \vert \theta \vert ^2 \pi^2 - 8\pi^4}{960 \pi^2 z^{\star 4}} + O(r^2)
\end{equation*}
The  divergent  $1/r^4$ term is simply characteristic of the ultra-violet conformal invariance of the theory \eqn{shorty}.  This should be renormalized through the addition of a suitable local counterterm. We define our regularization scheme to subtract this conformal piece. What remains is the expectation value of the composite operator $\langle\silly\rangle_{\rm reg}$. 
%
%
%\begin{equation}
%\langle \tfrac{1}{2} \bar{\Psi} ( \gamma^i \overrightarrow{\partial_i} - \overleftarrow{\partial_i} \gamma^i ) \Psi \rangle_{\text{reg}} = -\frac{15 \vert \theta \vert ^4 - 60 \vert \theta \vert ^3 \pi + 60 \vert \theta \vert ^2 \pi^2 - 8\pi^4}{960 \pi^2 z^{\star 4}} \label{derivcond}
%\end{equation}
%
Comparing to \eqn{goodf}, we can write
\be \langle\bar{\psi}\psi\rangle = \frac{1}{4\pi} - 2\langle\silly\rangle_{\rm reg}\, z^4+\ldots\nn\ee
%

%\begin{equation}
%\langle \tfrac{1}{2} \bar{\Psi} (\gamma^i \overrightarrow{\partial_i} - \overleftarrow{\partial_i} \gamma^i) \Psi \rangle_{\text{reg}} = - \tfrac{1}{2} \lim_{z \to 0} z^{-4} \langle \bar{\psi} \psi \rangle_{\text{sub}}
%\end{equation}

%The subscript on the RHS serves as a reminder to subtract away the $1 / 4 \pi^2$ term that exists even in full AdS before taking the limit $z \to 0$. It is akin to the subscript on the LHS, reminding us to subtract away the divergent, conformal part of the boundary two-point correlator before sending the two points together.

%\footnote{Once again, one may raise the objection as to whether regularisation will subtract away an extra finite piece from the condensate. But if this were the case, then by locality, $\langle \bar{\psi} \psi \rangle_{\text{full AdS}}$ and $\langle \bar{\psi} \psi \rangle_{\text{with wall}}$ would be afflicted by the same subtraction. Since $\langle \bar{\psi} \psi \rangle_{\text{sub}} = \langle \bar{\psi} \psi \rangle_{\text{with wall}}-\langle \bar{\psi} \psi \rangle_{\text{full AdS}}$, it would be unaffected even if such a subtraction were necessary.} 

This result tells us how to interpret the sub-leading fall-off of the $\bar{\psi}\psi$ condensate in the boundary theory. However, it does leave open the question of the meaning of the background constant condensate \eqn{4pi}. We don't have an answer to this; we suspect that it should simply be subtracted away when discussing the dictionary between bulk and boundary.

\subsubsection*{General Masses}

We now generalise this result to arbitrary bulk fermion mass. In what follows, we subtract the condensate in the full AdS geometry.\footnote{For $m \neq 0$, the condensate $\langle \bar{\psi}\psi\rangle$ in the full AdS geometry has a divergent piece that requires regularisation \cite{allen}. This does not contradict our earlier argument that the renormalization procedure does not subtract anything from $\langle \bar{\psi} \psi\rangle$ when $m = 0$. If $m \neq 0$, there is no chiral symmetry, so no relationship between the subtraction from $\langle i \bar\psi \Gamma^5 \psi \rangle$ and the subtraction from $\langle \bar\psi \psi \rangle$. }. What remains is finite and we denote it with a subscript ``sub" .
%
%
%\footnote{$\langle \bar{\psi}(z, \vec{x}) \psi(z, \vec{x}) \rangle_{\text{full AdS}} = z^4 \int \frac{d^3k}{(2\pi)^3}k   \left( 2K_{m+\frac{1}{2}}(kz) I_{m-\frac{1}{2}}(kz) - 2I_{m+\frac{1}{2}}(kz) K_{m-\frac{1}{2}}(kz) \right)$. It is constant in $z$, but for $m \neq 0$ it is divergent unless regularised \cite{allen}. 
%
\begin{equation}
\langle \bar{\psi}(z, \vec{x}) \psi(z, \vec{x}) \rangle_{\text{sub}} = z^4 \int \frac{d^3k}{(2\pi)^3}k \left( 4I_{m-\frac{1}{2}}(kz) I_{m+\frac{1}{2}}(kz) \right) {\cal M}_1 (k)
\end{equation} 
where ${\cal M}_1$ is again the Euclidean analog of \eqn{coeffs}, 
\begin{equation*}
{\cal M}_1 (k) \equiv \frac{-\cos^2(\tfrac{\theta}{2}) K_{m+\frac{1}{2}}(kz^\star)I_{m+\frac{1}{2}}(kz^\star) + \sin^2( \tfrac{\theta}{2}) K_{m-\frac{1}{2}}(kz^\star)I_{m-\frac{1}{2}}(kz^\star)} {\cos^2(\tfrac{\theta}{2}) I_{m+\frac{1}{2}}^2 (kz^\star) + \sin^2 (\tfrac{\theta}{2}) I_{m-\frac{1}{2}}^2(kz^\star)}.
\end{equation*}
Expanding near the boundary, this becomes
\be
\langle \bar{\psi} \psi \rangle_{\text{sub}} = \int \frac{d^3k}{(2\pi)^3} k \left( \frac{k}{2} \right)^{2m} \frac{4 {\cal M}_1 (k)}{(m + \tfrac{1}{2}) \Gamma(m + \tfrac{1}{2})^2} \, z^{4+2m}+\ldots
\nn\ee 
We would again like to relate this to the boundary expectation value. Comparing with \eqn{corr} and remembering to subtract off the ultra-violet conformal piece from the boundary two-point function, we obtain the result
\begin{equation}
\langle \bar{\psi} \psi \rangle_{\text{sub}}  = -\frac{2}{2m+1}\langle\silly\rangle_{\rm reg}\,z^{4+2m} +\ldots \label{genm}
\end{equation}
The subscript ``sub'' on the left is akin to the subscript ``reg'' on the right. When computing the bulk condensate, we subtracted away the contribution to the bulk Green's function from the full AdS geometry, leaving only the effect of the hard wall. Likewise, when computing the boundary condensate, we subtracted away the ultra-violet, conformal contribution to the boundary Green's function, leaving only the infra-red term due to the cut-off.

\subsubsection*{Aside: A sum over bounces}

The condensate \eqn{bouncey} in theory of massless fermions has a nice interpretation when Fourier decomposed in $\theta$ that highlights the method-of-images approach to bulk catalysis presented in \cite{magcat}. Let us define $\psi_L = \tfrac{1}{2}\left( 1 - \Gamma^5 \right) \psi$ and $\psi_R = \tfrac{1}{2}\left( 1 + \Gamma^5 \right) \psi$, so $\bar{\psi}_L \psi_R = \tfrac{1}{2}\bar{\psi} \psi + \tfrac{1}{2}\bar{\psi} \Gamma^5 \psi$. Then the condensate can be written as 
\be
\langle \bar{\psi}_L \psi_R \rangle = \frac{z^3}{8\pi^2}  \sum_{n = -\infty}^{\infty} \frac{e^{in \theta}}{(z + n z^\star)^3}. \nn\ee
This is the result of \cite{magcat}. It captures the physics of a massless fermion bouncing backwards and forwards between the UV boundary and the IR wall. The chirality of the fermion flips at each bounce. Moreover, it picks up a phase $e^{i\theta}$ each time it collides with the IR wall.  Since the chirality of the fermion flips at each bounce, only the trajectories containing an odd number of bounces contribute to $\langle \bar{\psi}_L \psi_R \rangle$. For $n \geq 0$, the $n^\text{th}$ term in the Fourier series represents propagation via $2n+1$ bounces, first bouncing off the UV boundary. For $n < 0$, the $n^\text{th}$ term in the Fourier series represents propagation via $-2n-1$ bounces, first bouncing off the IR wall. The full condensate is given by the sum over all possible trajectories. We will briefly revisit this interpretation in Section 5.

\section{Magnetic Catalysis and the Callan-Rubakov Effect}
\label{monwallsec}

In this section, we provide an alternative framework in which to discuss magnetic catalysis. Instead of focussing on the hard wall geometry, we will instead build an effective magnetic wall from bulk non-Abelian gauge fields and scalars. This solitonic domain wall is called the {\it monopole wall} and had been studied in both flat space \cite{bag,cherkis,ward} and in AdS \cite{monoads,sutcliffe}. 

\para
The monopole wall is not a simple object. In particular, it is not translationally invariant; the wall forms a lattice structure, woven from underlying 't Hooft-Polyakov monopoles. Nonetheless, in \cite{monoads}, a simple Abelian approximation of the monopole wall was developed (following \cite{bag}) which ignored these subtleties and allowed many of the relevant properties of the wall to be understood analytically. Subsequent numerical simulation showed this Abelian approximation to be in excellent agreement with the true solutions \cite{sutcliffe}.

\para
Like the magnetic hard wall (or, indeed, the magnetic \RN black hole), the long-range behaviour of the monopole wall corresponds to a constant, homogeneous magnetic field $B$ in the dual boundary theory. However, deep in the interior of AdS, it returns to the vacuum geometry. The phase transitions between the monopole wall and \RN black hole were explored in \cite{monoads}, where it was shown that for large regions of parameters the monopole wall is the preferred ground state. Moreover, the complicated lattice structure of the wall has interesting consequences for the boundary theory: it  corresponds to the dynamical formation of a crystal, 
 spontaneous breaking translational invariance.

\para
In this section, we study the dynamics of light fermions in the presence of the monopole wall. In flat space, it has been known for a long time that there is a beautiful interplay between fermions, anomalies and monopoles. In the presence of a background $B$  field, one needs only  a fluctuation of the electric field $E$ to generate a non-trivial $E\cdot B$. This enhances the the chiral anomaly, resulting in chiral symmetry breaking condensate surrounding the monopole,  without the accompanying exponential instanton suppression. The resulting phenomenology usually goes by the name of the Callan-Rubakov effect \cite{rubakov,callan1,callan2}.

\para
Below, we show that the Callan-Rubakov effect in the presence of the monopole wall is identical to the story of magnetic catalysis that we have seen in previous sections. We will argue that the monopole wall acts as a reflecting boundary for fermion in the lowest Landau level. In particular, the $\theta$-angle that parameterises the IR boundary conditions \eqn{sir} is identified with the usual CP-violating $\theta$-angle of non-Abelian gauge theories. The chiral symmetry breaking condensate in the bulk is identified with the magnetic catalysis condensate in the boundary.

\subsection{Monopoles in Global AdS}

Throughout this section, we will be interested in the dynamics of $SU(2)$ Yang-Mills theory, coupled to a single adjoint Higgs field $\phi$, in AdS space. The action is given by
\be S = -\int d^4x \sqrt{-g} \left[ 
\frac{1}{e^2} \left( \frac{1}{4}F^a_{\mu\nu}F^{a\mu\nu}+\frac{1}{2}{\cal D}_\mu \phi^a {\cal D}^\mu
\phi^a + V(\phi)   \right) + \frac{\theta}{16 \pi^2} \epsilon^{\mu\nu\rho\sigma} F^a_{\mu\nu}F_{a\mu\nu}\right]\ \ \ \ \ \label{yangmills}\ee
The potential $V(\phi)$ is given by
\be V(\phi) = \frac{\lambda}{8}(\phi^a\phi^a-v^2)^2 \ee
This induces a Higgs expectation value,  $\phi^a\phi^a=v^2$, breaking the $SU(2)$ gauge symmetry to $U(1)$. In the following, we assume that $vL\gg 1$, where $L$ is the AdS radius. 

\para
Under a CP transformation, $\theta\rightarrow -\theta$. This means that the $\theta$-angle breaks CP unless $\theta=0$ or $\theta=\pi$.

\para
In flat space, this action is well known to exhibit 't Hooft-Polyakov monopoles. However, like all solitons, their existence relies crucially on the boundary conditions for various fields and is correspondingly sensitive to the asymptotic nature of spacetime. Although we are ultimately interested in the Poincar\'e patch, we start by considering monopoles in global AdS. In Section \ref{wallsec} we will subsequently review the scaling limit under which monopoles in global AdS turn into monopole walls in the Poincar\'e patch. The metric on global AdS with unit radius, $L=1$, is
\be ds^2 = -\left(1+{r^2}\right)dt^2 +
\left(1+{r^2}\right)^{-1}dr^2 + r^2 d\Omega_2^2 \label{global}\ee
The conformal boundary, at $r\rightarrow \infty$, is  ${\bf R}\times {\bf S}^2$ which naturally allows a winding of the scalar expectation value on the ${\bf S}^2$, making use of the fact that $\Pi_2(SU(2)/U(1))\cong {\bf Z}$. 

\para
The solution for a single 't Hooft-Polyakov monopole takes the form
\be
A_i^a &=& \epsilon_{aij} \frac{\hat{r}_j}{r} A(r) \nonumber \\
A_0^a &=& 0 \nonumber \\
\phi^a &=&  \hat{r}_a v H(r)
\nn\ee
where the profile functions $H(r),A(r)$ depend on the AdS metric but the boundary conditions are the same as in flat space-time, they vanish at zero and saturate to one at infinity. These profile functions in AdS were studied in \cite{schap}. As in flat space, the monopole transforms covariantly under neither the rotation group $SU(2)_{\rm rot}$ nor the gauge group $SU(2)_{\rm gauge}$. Instead, the monopole locks these two groups together, with covariant transformation under the diagonal $SU(2)_{\rm diag}=SU(2)_{\rm rot}\times SU(2)_{\rm gauge}$.

\subsubsection*{Fermions and Monopoles}

The dynamics of light fermions in the presence of a monopole exhibit a number of interesting properties, first explored in  \cite{jackiw}. Since monopoles in global AdS  are very similar to those in flat space, much of the discussion in the original literature regarding fermions in this background can be immediately transplanted to the present context. 

\para
We add to the action \eqn{yangmills} a single Dirac fermion, $\psi$ transforming in the fundamental representation of $SU(2)_{\rm gauge}$. Its action is given by \eqn{sbulk}, where the covariant derivative now conceals both spin and gauge connections. The discussion of the Callan-Rubakov effect is simplest if we restrict to a massless fermion \cite{rubakov,callan1}, although the addition of a mass does not change the main conclusions \cite{callan2}; in the remainder of this section we set $m=0$.

\para
The relevant physics takes place in the s-wave  of $SU(2)_{\rm diag}$ and, for this reason, the problem can be reduced to the half-line parameterised by $r\in[0,\infty)$. For fermions, the locking of gauge and rotation symmetries results in a correlation between the charge and four-dimensional helicity for s-waves. (This is related to the fact that the lowest relativistic Landau level carries half the states of the higher levels).  A single fundamental Dirac fermion in four dimensions decomposes into two chiral fermions on the half-line, twice the number of states that we had in the previous sections due to the extra gauge index. The left-moving fermion, $\xi_-$, gives rise to scattering states moving towards the monopole; the right-moving fermion $\xi_+$ gives rise to states scattering away from the monopole. The charge and four-dimensional helicity of each of these states is
\be
\begin{array}{cccc}
\quad {\rm Field} \quad  & \quad {\rm Direction} \quad& \quad {\rm Charge}  &  \quad{\rm Helicity}\quad \\
\xi_- & {\rm IN} &    + & L \\
\bar{\xi}_- & {\rm IN} & - & R\\
\xi_+ & {\rm OUT} & - & L \\
\bar{\xi}_+ & {\rm OUT} & + & R
\nn\end{array}
\ee
As an incoming fermion scatters off the monopole, it must necessarily change either its charge of its helicity. The naive guess would be that the fermion flips its charge, exciting the monopole into a dyon in the process. In fact this cannot happen for low-energy fermions since there is an energy gap, proportional to the W-boson mass, to the creation of a dyon. (For a particularly clean discussion of this process, see \cite{polchinski}). Instead, the fermion must bounce off the monopole, preserving its charge but flipping its helicity in the process. The resulting boundary conditions break chiral symmetry in the core of the monopole \cite{rubakov,callan1}.

\para
There are two further ingredients that we require. The first concerns the finite size of the monopole, $r^\star\approx 1/v$. Although much of the early literature on the Callan-Rubakov effect treats the monopole as effectively point-like, a more careful analysis shows that the incoming fermions are exponentially suppressed in the core of the monopole: a better approximation is to treat the bounce as occurring at $r=r^\star$ \cite{big,bigger}. (This has little effect for the Callan-Rubakov effect in flat space, but will prove important for our story). 

\para
The final ingredient is the role of the $\theta$-angle. The axial anomaly allows us to absorb the $\theta$-angle by a rotation of the chiral phase of the four-dimensional fermion. The net result is that the boundary condition that should be imposed on the s-wave fermion at the monopole is 
\be
\label{chiralbreaking}
\xi_+(r^\star) = e^{i \theta} \xi_-(r^\star)
\ee
These are equivalent to the one-parameter family of boundary conditions introduced in \eqn{sir} through the use of a hard wall. 
%
% \FIGURE{\epsfig{file=twopossibilities.eps,width=15cm} 
 %       \caption[]{Two possible scattering processes for the s-wave bouncing off the monopole. The first preserves charge and helicity, but is not energetically allowed; the second is the actual scattering process. It violates helicity. }
%	\label{twopossibilities}}
%

\para
The boundary conditions \eqn{chiralbreaking} induce a chiral condensate  close to the monopole which is suppressed by neither $1/v$ nor the usual $e^{-1/e^2}$ instanton factor: this is the Callan-Rubakov effect. At distances $r\gg r^\star$, but still much smaller than the AdS scale, the s-wave condensate $\langle\bar{\xi}_+\xi_-\rangle$ translates into the usual flat-space condensate for the four-dimensional fermions, 
\be  \langle\bar{\psi}\psi\rangle + \langle\bar{\psi}\Gamma^5\psi\rangle  = \frac{e^{i\theta}}{8\pi^2 r^3}\nn\ee

In AdS, there is one further complication: massless fermions can reach the boundary at $r=\infty$ in finite time. To complete the story, we must also impose suitable boundary conditions at the boundary. The appropriate boundary conditions are those inherited from \eqn{suv} and these also break chiral symmetry. 
 The resulting picture is of a fermion bouncing backwards and forwards between monopole and boundary, flipping its helicity after each bounce. This propagator, and the condensate is produces, was the subject of \cite{magcat}. Rather than reviewing this story, let us first explain how things change as we move from the monopole in global AdS to the monopole wall in the Poincar\'e patch.

\subsection{The Monopole Wall}
\label{wallsec}

The Poincar\'e patch of AdS has asymptotic boundary ${\bf R}^{1,2}$ which does not appear to allow for the winding of a Higgs field necessary to support monopole solution.  In this context, the monopole wall is more natural. One can construct the monopole wall from the 't Hooft-Polyakov monopoles by taking a suitable scaling limit, zooming in on the Poincar\'e patch while simultaneously cranking up the monopole charge \cite{monoads}. This is entirely analogous to the way that one constructs  the planar black hole in the Poincar\'e patch from the large black holes in global AdS. Under this scaling, the winding of the Higgs field is hidden behind the Poincar\'e horizon; all that remains of the boundary conditions is a constant magnetic field $B$ for the $U(1)\subset SU(2)_{\rm gauge}$ on the boundary.

\para
While the full solution to the monopole wall in AdS is complicated and known only numerically \cite{sutcliffe}, progress can be made by focussing on the Abelian $U(1)$ fields far from the wall \cite{bag}, subject to certain boundary conditions at the location of the wall, $z=\zstar$. If we take the particularly simple case of vanishing potential, $\lambda=0$, then this Abelian solution  for $z\leq \zstar$ is given by
\be
\phi = v\left(1-\frac{z^3}{z^{\star\,3}}\right) \ \ \ \ ,\ \ \ \ B={\rm constant}\nn\ee
In contrast, in the IR of the geometry $z>\zstar$, all fields vanish and the full $SU(2)$ symmetry is restored. The position of the wall is determined dynamically by the strength of the magnetic field \cite{monoads},
\be \zstar = \sqrt{\frac{3v}{ B}}\label{r0}\ee
A non-vanishing potential introduces some mild $\lambda$ dependence into this formula\footnote{In flat space, there are no multi-monopole solutions, and hence no monopole wall, when $\lambda>0$. Only when $\lambda=0$, and the Higgs field is massless, is the  magnetic repulsion of the monopoles exactly cancelled by the scalar attraction. However, in AdS this restriction is not necessary since the magnetic repulsion can be balanced against gravitational attraction.. For $\lambda\approx 0$, the multi-monopole solutions in AdS are expected to be close to their BPS counterparts in flat space which can be described by the bag approximation \cite{bag}. However, as $\lambda$ increases, the monopoles become more point like and we expect a transition to something more akin to a Wigner crystal of Dirac-like monopoles}.

\para
The wall spits out magnetic field in the UV region $z<\zstar$ which stratifies any fermions into Landau levels. The s-wave fermions in global AdS become the lowest Landau level in the Poincar\'e patch; each have half the states of higher modes. (The relationship between $\xi_\pm$ and the four-component Dirac fermions $\psi$ in in the magnetic wall is given in \cite{magcat}). We claim that the lowest Landau level fermions inherit the boundary condition \eqn{chiralbreaking} at the location of the monopole wall,
\be \xi_+(\zstar)=e^{i\theta}\xi_-(\zstar)\nn\ee
These should be supplemented with the boundary conditions at $z=0$ arising from \eqn{suv}.

\para
With these boundary conditions in place, the remaining calculation is identical to that of \cite{magcat}, albeit with an additional factor of two coming from the implicit sum over gauge indices in expressions such as $\bar{\psi}\psi$ and $\bar{\xi}_+\xi_-$. The propagator $\langle\bar{\xi}_+\xi_-\rangle$ involves an infinite number of bounces between the monopole wall and boundary and can be written as,
\be    \langle\bar{\psi}\psi\rangle + \langle\bar{\psi}\Gamma^5\psi\rangle = \frac{B}{\pi} \langle\bar{\xi}_+\xi_-\rangle = -\frac{Be^{i\theta}}{4\pi^2}\sum_{n=-\infty}^{+\infty}\frac{e^{in\theta}z^3}{z+n\zstar}\nn\ee
Of particular interest is the behaviour as $\theta\rightarrow 0$. As explained in \cite{magcat}, an infra-red divergence ensures that the sum is not continuous in this limit and, despite appearances, is not real. The net result is appearance of a bulk condensate, spontaneously breaking CP invariance
\be \lim_{\theta\rightarrow 0} \ \langle i\bar{\psi}\Gamma^5\psi\rangle = \pm \frac{1}{4\pi}\frac{B}{z\star}\,z^3.\label{donenow}\ee

\subsection{A Final View from the Boundary}

The translation of the bulk condensate \eqn{donenow} into the boundary is simply done using the results of Section 4. The relationship \eqn{ilikethis} tells us that the bulk condensate is dual to a boundary condensate $\langle\bar{\Psi}\Psi\rangle$. Indeed, the only novelty compared to  the hard wall calculation \eqn{lovely} is that the position of the monopole wall $\zstar$ itself depends on the strength of the magnetic field \eqn{r0}. This results in a different scaling of the condensate with $B$. 
\be
\lim_{\theta \to 0}\  \langle \bar{\Psi} \Psi \rangle =\pm \frac{B^{3/2}}{4 \pi \sqrt{3 v}} \ {\rm sign}(\theta)
\ee
The upshot is that, when placed in the context of AdS/CFT, the well studied Callan-Rubakov effect in the bulk gives rise to magnetic catalysis in the boundary. The characteristic $B^{3/2}$ scaling distinguishes this process from catalysis arising from the fixed hard wall geometry, in which the boundary condensate grows linearly with $B$.

%\para
%We previously considered the CR effect treating the monopole as a point like object which just define the particular bouncing condition for the light fermions.
%We have not used the condition $v \ll M_{Pl}$ explicitly since all the dynamics we are interested in is at a much lower scale than $v$.  but in fact we are using this condition at an important step, when we say that the fermion actually is bouncing back of the monopole in finite time.
%When $v$ reaches the Plank mass and the monopole becomes a magnetic black-hole, every particle falling inside the hole does not bounce back. We thus loose the CR effect and no chiral condensate is formed around the magnetic black hole.  Thus the CR effect is a distinctive feature of the soliton phase with respect to the black hole phase of the monopole. We could wonder where the $\dot{Q}_5 \propto \int \vec{E} \vec{B}$ argument fails in the case of a dyonic black hole. The reason is the well known fact that in gravity global symmetries do not exist, loosely speaking. the chiral charge is in fact continuously flowing inside the dyonic black hole and no remnant of it remain from outside.

\appendix
\section{Appendix: Doubling up in AdS$_{5}$}
\label{app5}

In this appendix, we present a short summary of the most important results of Sections 2,3 and 4 applied to spinors in AdS${}_5$. A 4-component Dirac spinor in AdS${}_5$ is dual to a two-component Weyl spinor in the $d=3+1$ dimensional boundary theory. To construct a Dirac spinor in the boundary theory, we require two bulk fermions, $\psi$ and $\chi$ which we take to have equal and opposite masses, $m$ and $-m$. 

\para
The boundary Dirac spinor $\Psi$ naturally decomposes into left- and right- handed Weyl spinors, $\Psi_L$ and $\Psi_R$.  The bulk spinor $\psi^{-}$ acts as a source for $\Psi_L$ while $\chi^+$ acts as a source for  $\Psi_R^{\dagger}$. (The equal and opposite masses ensure that both sources have the same fall-off in the UV). 

%
%\be \Psi= \left( \begin{array}{c} \Lambda \\ \Sigma^{\dagger} \end{array} \right) \qquad {\rm and} \qquad \bar{\Psi}= \left( \begin{array}{cc} \Sigma & \Lambda^{\dagger} \end{array} \right)\ .\nn\ee
%
%The Weyl spinor  $\Lambda$ is sourced by $\psi^{-}$ while $\Sigma^{\dagger}$ is sourced by $\chi^{+}$. (The equal and opposite masses ensure that both sources have the same fall-off in the UV). 

%
% \begin{equation*}
%\exp \left( i S_{\text{AdS}}(\psi_0^-, \bar{\psi}_0^-, \chi_0^+, \bar{\chi}_0^+) \right)  = \exp \langle i \int \left( \Lambda^\dagger \psi_0^- + \bar{\psi}_0^- \Lambda + \Sigma \chi_0^+ + \bar{\chi}_0^+ \Sigma^\dagger \right) \rangle
%\end{equation*}
%
%A convenient gamma matrix representation is 
%
%\begin{equation}
%\Gamma^{z} = \left( \begin{array}{cc} 1 & 0 \\ 0 & -1 \end{array} \right) \qquad
%\Gamma^{0} = \left( \begin{array}{cc} 0 & -i \\ -i & 0 \end{array} \right) \qquad
%\Gamma^{i} = \left( \begin{array}{cc} 0 & -i\sigma^{i} \\ i\sigma^{i} & 0 \end{array} \right) \quad (i=1,2,3).
%\end{equation}

\para
Although there is no chiral symmetry in AdS${}_5$, we can still impose a one-parameter family of boundary conditions on the IR wall by mixing up the $\psi$ and $\chi$ fields. The IR action  is taken to be
\begin{equation}
S_{\text{IR}}  = \tfrac{1}{2} \int _{z = z^{\star}}  d^{4}x \sqrt{-h} \left( \cos \theta (\bar{\psi} \psi - \bar{\chi} \chi) - \sin \theta (\bar{\psi} \chi + \bar{\chi} \psi) \right)
\end{equation}
which imposes the IR boundary condition 
\begin{equation}
\cos \left(\tfrac{\theta}{2}\right)\, \psi^{-} - \sin \left(\tfrac{\theta}{2}\right)\, \chi^{-} = 0, \qquad \sin (\tfrac{\theta}{2}) \psi^{+} + \cos( \tfrac{\theta}{2} )\chi^{+} = 0.\nn
\end{equation}
All the technology of Section~\ref{sec:2} duly transfers to AdS$_5$. For example, the AdS$_5$ spectrum is identical to that of  AdS$_4$,
\begin{equation}
\frac{J_{m+\frac{1}{2}}(kz^{\star})}{J_{m-\frac{1}{2}}(kz^{\star})}= \pm \tan \tfrac{\theta}{2}.
\end{equation}
The AdS$_5$ boundary two-point function also mirrors that of AdS$_4$ \eqref{corr}. 
\begin{equation}
\langle \bar{\Psi}_{\beta}(\vec{y}) \Psi_{\alpha}(\vec{x}) \rangle \\
= -i \int \frac{d^{4}k}{(2\pi)^{4}} e^{-i \vec{k}.(\vec{x}-\vec{y})} \left( \frac{k}{2} \right)^{2m}  \left( - i \vec{\Gamma}_{\alpha \beta}.\hat{\vec{k}} N_{1}(k) + \delta_{\alpha \beta} N_{2}(k) \vphantom{\hat{\vec{k}}} \right)\nn
\end{equation} 
In the presence of a magnetic hard wall, we can repeat the calculations of Section \ref{sec:4}. However, the dimensionality of the spacetime turns out to be all-important.  The magnetic field once again reduces the effective dimensionality by two, so this time the lowest Landau level dynamics takes place in $d=1+1$ dimension. Correspondingly, the integral \eqn{inttodo} becomes 2-dimensional. 
Its behaviour as $\theta \to 0$ is given by 
\be
\lim_{\theta \to 0^\pm}\ \langle \bar{\Psi} \Psi \rangle_{\text{AdS}_5} = \lim_{\theta \to 0^\pm} \frac{B}{\pi z^{\star 2m}} \int_0^\Lambda \frac{dk}{2 \pi}\frac{\theta k}{\left( \frac{kz^\star}{2m+1} \right)^2 + \left( \frac{\theta}{2} \right)^2} \nn\ee
We have introduced an ultraviolet cutoff, $\Lambda$, because the integrand is only valid for small $k$: we have expanded in $k$ to lowest order. In the limit $\theta\rightarrow 0$, the integrand vanishes at high $k$. Moreover, as expected, the extra dimension softens the infrared divergence and for small $\theta$ the condensate behaves as $\theta\log\theta$,
\be
\lim_{\theta \to 0^\pm}\ \langle \bar{\Psi} \Psi \rangle_{\text{AdS}_5} 
 = \lim_{\theta \to 0^\pm} \frac{B(2m+1)^2}{4\pi^2 z^{\star 2m+2}}\quad \theta \log \left( 1 + \left( \frac{\Lambda z^{\star}}{(m+\tfrac{1}{2}) \theta} \right)^2 \right) = 0.
\nn\ee
This is similar to the behaviour seen in weak coupling calculations \cite{magcat1}. 

\para
Finally, despite the lack of magnetic catalysis, we may still search for the dictionary between composite operators outlined for in Section \ref{sec:3}. The analogous results in AdS${}_5$ take the form,
\begin{eqnarray}
\langle \bar{\Psi} \Psi \rangle  =  - 2 \lim_{z \to 0} z^{-4-2m} \langle \bar{\psi} \chi + \bar{\chi} \psi \rangle \nn\ee
and 
\be
\langle \bar{\Psi}\, \Gamma^i \!\stackrel{\leftrightarrow}{\partial_i} \!\Psi \rangle_{\text{reg}}  =  -(2m+1) \lim_{z \to 0} z^{-5-2m} \langle \bar{\psi} \psi - \bar{\chi} \chi \rangle_{\text{sub}}\nn
\end{eqnarray}

\section{Appendix: Fermions in the Magnetic Hard Wall}
\label{magwallapp}

The computation of the two-point function in the magnetic hard wall background follows closely the algebra of Section \ref{sec:2}: $k$ becomes $q_n(\omega)$ and $i \Gamma . \hat{\vec{k}}$ becomes $\frac{i \omega}{q_n(\omega)}\Gamma^t + \frac{i \sqrt{2Bn}}{q_n(\omega)}\Gamma^y$, while $e^{-i k_x x}$ becomes $X_{n-1}(k,x) Q_{(+)} + X_{n}(k,x) Q_{(-)}$.

\para
The IR boundary conditions \eqref{wallbc} relates the $b_n (\vec{k})$ modes to the $a_n (\vec{k})$ modes in the expansion \eqn{landaulev}.
 \begin{multline}
 \left( b_{n(-)}^{-}(k,\omega) + b_{n(+)}^{-}(k,\omega) \right) = \left[ M_{1}(q_{n}(\omega))+\left( \frac{\omega}{q_{n}(\omega)} \Gamma^{5} \Gamma^{t} + \frac{\sqrt{2Bn}}{q_{n}(\omega)} \Gamma^{5} \Gamma^{y} \right) M_{2}(q_{n}(\omega)) \right] \\
\times (a_{n(-)}^{-}(k,\omega) + a_{n(+)}^{-}(k,\omega))\nn
\end{multline} 
where the functions $M_1$ and $M_2$ are defined in \eqn{coeffs}. It follows that
 \begin{multline}
 \left( b_{n(-)}^{+}(k,\omega) + b_{n(+)}^{+}(k,\omega) \right) = \left[ - \left( \frac{i\omega}{q_{n}(\omega)} \Gamma^{t} + \frac{i \sqrt{2Bn}}{q_{n}(\omega)} \Gamma^{y} \right) M_{1}(q_{n}(\omega)) - i \Gamma^5 M_{2}(q_{n}(\omega)) \right]\\
\times (a_{n(-)}^{-}(k,\omega) + a_{n(+)}^{-}(k,\omega)).\nn
\end{multline}
The spectrum can be extracted from the pole structure of $M_{1}(q_{n}(\omega))$ and $M_{2}(q_{n}(\omega))$. It is given by solutions to 
\begin{equation}
\frac{J_{m+\frac{1}{2}}(z^{\star} \sqrt{\omega^{2}-2Bn} )}{J_{m-\frac{1}{2}}(z^{\star} \sqrt{\omega^{2}-2Bn})} = \pm \tan \tfrac{\theta}{2}
\end{equation} 
As we described in the main text, when $\theta=0$, the lowest Landau level $n=0$ contains zero modes, indexed by $k$.  None of these modes individually respect translational symmetry in $y$, but that translational symmetry is restored when we integrate over $k$. 
\begin{equation*}
\psi_{\omega=0, n=0} (z, x, y, t) = \int \frac{dk}{2\pi} e^{-iky} z^{\frac{3}{2}+m} X_{0}(k,x) b_{0(-)}^{+}(k)
\end{equation*}
Retracing the steps of Section \ref{sub:2.3}, we can compute the on-shell AdS action. Expressing the responses $\psi_{0(\pm)}^{+} = \lim_{z \to 0} z^{-(\frac{3}{2}+m)} \psi_{(\pm)}^{+}$ in terms of the sources $\psi_{0(\pm)}^{-} = \lim_{z \to 0} z^{-(\frac{3}{2}-m)} \psi_{(\pm)}^{-}$, the on-shell action  evaluates to 
\begin{multline}
S_{\text{AdS}} = - \int dx dy dt dx' dy' dt' \frac{dk d\omega}{(2\pi)^{2}} \sum_{n} e^{-i\omega(t-t')-ik(y-y')} \left( \frac{q_{n}(\omega)}{2} \right)^{2m} \\
\times \left( X_{n-1} (x,k) \bar{\psi}_{0(+)}^{-}(x,y,t) + X_{n} (x,k) \bar{\psi}_{0(-)}^{-} (x,y,t) \right) \\ \qquad \qquad \quad
\times \left( \left( \frac{i \omega}{q_{n}(\omega)}\Gamma^{t} + \frac{i \sqrt{2Bn}}{q_{n}(\omega)}\Gamma^{y} \right) N_{1} (q_{n}(\omega)) + i \Gamma^{5} N_{2}(q_{n}(\omega)) \right) \\
\times \left( X_{n-1}(x',k) \psi_{0(+)}^{-}(x',y',t') + X_{n} (x',k) \psi_{0(-)}^{-}(x',y',t') \right)\nn
\end{multline}
 In the representation \eqref{rep}, the components of $\psi$ are $\psi^{T} = (\psi_{(-)}^{+}, \psi_{(+)}^{+}, \psi_{(-)}^{-}, \psi_{(+)}^{-})$. One arrives at the boundary two-point correlator by taking functional derivatives of the action with respect to the sources $\psi_{0(\pm)}^{-}$. In fact, we will focus our labour on the trace part of the correlator with a view to computing the CP-violating condensate $\langle \bar{\Psi} \Psi \rangle$. We have
 \begin{eqnarray}
\langle\bar{\Psi}(0,0,0) \Psi(x,y,t) \rangle & = & -i \int \frac{dk d\omega}{(2\pi)^2} \sum_{n} \left( \frac{q_{n}(\omega)}{2} \right)^{2m} e^{-i\omega t - iky} \notag \\
&& \quad \times \left(\left(X_{n-1}(0,k)X_{n-1}(x,k) - X_{n}(0,k)X_{n}(x,k) \right)\frac{\omega}{q_{n}(\omega)} N_{1}(q_{n}(\omega)) \right. \notag \\
 && \qquad \left. + \left(X_{n-1}(0,k)X_{n-1}(x,k) + X_{n}(0,k)X_{n}(x,k) \right) N_{2}(q_{n}(\omega))  \vphantom{\frac{\omega}{q_{n}}}  \right)\nn
\end{eqnarray}

\section{Appendix: Motivating the Correspondence between Bulk and Boundary Condensates}
\label{hideapp}

We showed in Section 4 how the boundary two-point function for fermions are captured by the bulk two-point functions. In this Appendix, we will expose the underlying reasons behind this correspondence. Although we perform the calculations explicitly in the AdS hard wall background, a similar computation holds in any asymptotically AdS geometry. The first step is to show that, for a bulk fermion of mass $m$, the boundary two point function can be written as
\begin{equation}
\langle \bar{\Psi}_\beta (\vec{y}) \Psi_\alpha (\vec{x}) \rangle = -  \lim_{z \to 0} \ \frac{1}{2z^{3+2m}} \left( \vec{\gamma}_{\alpha \beta} \cdot \langle \bar{\psi}(z, \vec{y})\, \vec{\Gamma} \,\psi (z, \vec{x}) \rangle + \delta _{\alpha \beta} \langle i \bar{\psi}(z, \vec{y}) \Gamma^5 \psi(z, \vec{x})\rangle \right)\label{bulkbdy}
\end{equation} 

\para
This result is a manifestation of the fact that the bulk-to-boundary propagator $\Sigma(z, \vec{x} - \vec{y})$ is the limit of the bulk-to-bulk propagator $G(z, \vec{x}, w, \vec{y})$ as the point $(w, \vec{y})$ is taken towards the UV boundary. By definition, $\Sigma(z, \vec{x} - \vec{y})$  constructs the full classical solution $\psi(z, \vec{x})$ from the  source $\psi_{0}^{-} (\vec{x})$, 
 \begin{equation*}
\psi(z,\vec{x}) \equiv \int d^{3}y \ \Sigma(z, \vec{x}-\vec{y})\, \psi_{0}^{-}(\vec{y})
\end{equation*}
In the AdS hard wall background, the bulk-to-boundary propagator takes the form
\begin{equation}
\Sigma(z, \vec{x} - \vec{y} ) = - \frac{\pi}{\Gamma(m+\tfrac{1}{2})} \int \frac{d^{3} k}{(2\pi)^{3}} e^{-i \vec{k} . (\vec{x} - \vec{y})} \left( \frac{k}{2} \right) ^ {m+\frac{1}{2}} \phi_{IR} (z, \vec{k})\nn
\end{equation}
A little algebra confirms that this is indeed the purported limit of the bulk-to-bulk propagator  \cite{muck,kawano}.
\begin{equation}
\lim_{w \to 0} w^{-(\frac{3}{2} + m)}G(z,\vec{x}, w, \vec{y}) = - \Sigma(z, \vec{x}-\vec{y}) P^{-}\nn
\end{equation} 
It follows that the response $\psi_{0}^{+}$ to a source $\psi_{0}^{-}$ can be written as
\begin{equation}
\psi_{0}^{+}(\vec{x}) = - \lim_{z \to 0} \lim_{w \to 0} \int d^{3}y\ (zw)^{-(\frac{3}{2}+m)} \,P^{+} G(z, \vec{x}, w, \vec{y}) \psi_{0}^{-}(\vec{y})\nn
\end{equation} 
By the hermiticity property $\Gamma^0G(z, \vec{x}, w, \vec{y})^\dagger\Gamma^0=-G(w, \vec{y}, z, \vec{x})$, the on-shell boundary action \eqn{chocolate} is 
\begin{equation}
S_{\text{AdS}} = \lim_{z \to 0} \lim_{w \to 0} \int d^{3}x d^{3}y \ (zw)^{-(\frac{3}{2}+m)}\, \bar{\psi}_{0}^{-}(\vec{x})G(z , \vec{x}, w, \vec{y}) \psi_{0}^{-} (\vec{y}).\nn
\end{equation} 
Comparing this  with the on-shell action \eqref{onshell} computed in Section~\ref{sub:2.3}, we obtain 
\begin{multline}
\lim_{z,w \to 0}  (zw)^{-(\frac{3}{2}+m)} P^{+} G(z, \vec{x}, w, \vec{y}) P^{-} \\ = - \int \frac{d^{3}k}{(2\pi)^{3}} \left( \frac{k}{2} \right)^{2m} e^{-i \vec{k} \cdot (\vec{x} - \vec{y})}P^{+} \left( i \vec{\Gamma} \cdot \hat{\vec{k}} N_{1} (k) + i \Gamma^{5} N_{2} (k)  \right) P^{-}.\nn
\end{multline}
This is almost the result \eqn{bulkbdy} we set out to prove: all that remains is to make sense of the projection operators $P^\pm$. On the left-hand-side, these project onto the $Y_{m-\frac{1}{2}}$ and $J_{m-\frac{1}{2}}$ Bessel functions. These encode the $\psi^{+}$ character with the $z^{\frac{3}{2}+m}$ fall-offs required to attain the correct conformal dimension for the boundary operators. Meanwhile, the Bessel functions $Y_{m+\frac{1}{2}}(kz)$ and $J_{m+\frac{1}{2}}(kz)$ that encode $\psi^{-}$ character only appear in the products $Y_{m+\frac{1}{2}}(kz) J_{m+\frac{1}{2}}(kz)$ and $J_{m+\frac{1}{2}}^{2}(kz)$ in the expressions for $\text{tr}\left(G(z, \vec{x}, w, \vec{y})\Gamma^{i} \right)$ and $\text{tr} \left(G(z, \vec{x}, w, \vec{y}) \Gamma^{5} \right)$, and neither product contributes at highest order because $J_{m+\frac{1}{2}}(kz)$ has no leading series.

\para
Multiplying by $\Gamma^i$ or $\Gamma^5$ on both sides and taking the trace over bulk spinor indices, we obtain a relation \eqn{bulkbdy} between bulk and boundary correlators as promised. 

\para
Sending $\vec{y}\to \vec{x}$ and taking the trace over boundary spinor indices, we recover the correspondence \eqn{genm5} that relates $\bar\psi \Gamma^5 \psi$ to $\bar\Psi \Psi$.

\subsubsection*{The $\bar{\psi}\psi$ Condensate}

To motivate the relationship \eqn{genm}, we start by manipulating the classical Dirac equation \eqn{dirac}. We write
\be
 \frac{1}{2z^{3+2m}}&\bar{\psi}(z,\vec{y})&\!(\Gamma^{i}\! \stackrel{\rightarrow}{\partial}_{x^{i}} - \stackrel{\leftarrow}{\partial}_{y^i}\!\Gamma^{i} )\psi (z,\vec{x}) \notag \\
&= &  \frac{1}{z^{3+2m}} \left( \bar{\psi}^{+} (z,\vec{y})\!\stackrel{\leftrightarrow}{\partial}_z\!\psi^{-}(z,\vec{x}) - \bar{\psi}^{-}(z,\vec{y}) \!\stackrel{\leftrightarrow}{\partial}_z\! \psi^{+}(z,\vec{x}) \right) \notag \\
&& \qquad  + \frac{m}{z^{4+2m}} \Big( \bar{\psi}^{-}(z,\vec{y}) \psi^{+}(z,\vec{x}) + \bar{\psi}^{+}(z,\vec{y}) \psi^{-}(z,\vec{x}) \Big)\nn
\ee
We must deal with the leading and sub-leading parts of $\psi^+$ and $\psi^-$ separately. The contribution to the RHS from taking the leading parts of both $\psi^{+}$ and $\psi^{-}$ is zero. The next highest contribution, which comes from taking the leading fall-off for $\psi^{+}$ and pairing it with the sub-leading fall-off for $\psi^{-}$, has $z^{4+2m}$ asymptotics, and is the contribution we are interested in. This then allows us to relate $\bar{\psi}\psi$ to $\bar{\psi}\Gamma^i\psi$, 
\be
\lim_{z\to 0}\ \frac{1}{2} \frac{1}{z^{3+2m}}\bar{\psi}(z,\vec{y})(\Gamma^{i}\! \stackrel{\rightarrow}{\partial}_{x^{i}} - \stackrel{\leftarrow}{\partial}_{y^i}\!\Gamma^{i} )\psi (z,\vec{x}) \sim \frac{m+\ft12}{z^{4+2m}}\, \bar{\psi}(z,\vec{y}) \psi(z,\vec{x})\nn
\ee
Now we sandwich this classical equation of motion between quantum bras and kets and read off the relevant information from \eqref{bulkbdy}. It is tempting to send $\vec{y} \to \vec{x}$ prematurely and write down \eqref{genm} straight away. But there is one last subtlety to attend to. As things stand, both sides of the equation diverge in the limit $\vec{y} \to \vec{x}$. We must subtract from the LHS the contribution to $\langle \bar{\psi}(z, \vec{y}) \psi (z, \vec{x}) \rangle$ that exists in full AdS, and correspondingly we must also regularise away the conformal term from the RHS. Only after these subtractions have been made are the two-point correlators finite and continuous at $\vec{x} = \vec{y}$; it is only at this point that we entitled to switch the order of the limits $z \to 0$ and $\vec{y} \to \vec{x}$ and identify the condensates of the two composite operators. Thus we arrive at \eqref{genm}.

\section*{Acknowledgements}

We're grateful to Ofer Aharony, Micha Berkooz, Mike Blake, Aleksey Cherman, Sungjay Lee and Shimon Yankielowicz for useful discussions. S.B. is supported by the Lady Davis fellowship of the University of Jerusalem.  J.L., D.T. and K.W. are supported by the ERC STG grant 279943, ``Strongly Coupled Systems".

\end{document}